\documentclass[letterpaper, 10 pt, journal]{IEEEtran}

\usepackage[utf8]{inputenc} % allow utf-8 input
\usepackage[T1]{fontenc}    % use 8-bit T1 fonts
\usepackage{hyperref}       % hyperlinks
\usepackage{url}            % simple URL typesetting
\usepackage{booktabs}       % professional-quality tables
\usepackage{amsfonts}       % blackboard math symbols
\usepackage{nicefrac}       % compact symbols for 1/2, etc.
\usepackage{microtype}      % microtypography
\usepackage{moreverb,url}

\usepackage{hyperref}
\usepackage{amsmath} % allows for mathematical symbols
\usepackage{amssymb} % defines symbol names for all the math symbols
\usepackage{graphicx} %   for .pdf graphics inclusion
\graphicspath{{figures/}{figures/}}
\usepackage{subfig}
\usepackage{caption}
\usepackage[calcwidth = \columnwidth]{caption}
\usepackage{booktabs,threeparttable}	
\usepackage{setspace} %	change the spacing inside a document
\usepackage{mathptmx}
\usepackage{textcomp}
\usepackage{float} %	improves the interface for defining floating objects
\usepackage{rotating}
\usepackage{wrapfig}
\usepackage{color}
\usepackage{algorithm}
\usepackage[noend]{algpseudocode}
\usepackage{multirow}
\usepackage{eqnarray}
\usepackage{hhline}

\usepackage[symbol]{footmisc}

\usepackage{epsfig} % for postscript graphics files
\usepackage{mathtools}
\usepackage{tabularx}
\usepackage{array}% http://ctan.org/pkg/array

\usepackage{algorithm}
\usepackage{algpseudocode}
\usepackage{amsmath,amsfonts,amssymb}
\usepackage{array}

\begin{document}

\title{\LARGE \bf
Parameterized and GPU-Parallelized Real-Time Model Predictive Control for High Degree of Freedom Robots}

\author{Phillip Hyatt$^1$\: Connor S. Williams$^1$\: Marc D. Killpack$^1$

\thanks{(1) Authors are affiliated with the Mechanical Engineering department at Brigham Young University, Provo, UT, 84602 USA}}

\markboth{Pre-print Paper, Submitted January 2020}%
{}

\maketitle

\begin{abstract}
%A method for solving high Degree of Freedom (DoF) Model Predictive Control (MPC) problems at real-time rates is presented which makes use of the Graphics Processing Unit (GPU) available in most computers. This form of MPC uses an evolutionary algorithm and is called Evolutionary MPC (EMPC). Through simulation and hardware experiments, EMPC is shown to perform similarly to conventional MPC methods for low DoF cases and is shown to have faster solve times and (hopefully) therefore better performance than conventional MPC for high DoF cases. These advantages are likely to become more dramatic with the rise of more powerful and efficient GPUs. Hardware experiments also demonstrate that EMPC solve times are less affected by estimation and message passing processes which compete with conventional MPC for CPU time.
%Because EMPC does not improve upon an existing or learned policy it is less susceptible to falling in local minima. This is made possible in part by parameterizing the search space of control inputs and searching over the smaller space of parameters. An experimental study is presented which demonstrates the effects of several parameterizations on MPC performance. Model Predictive Control (MPC) is a form of optimal control used frequently in robotics. 
This work presents and evaluates a novel input parameterization method which improves the tractability of model predictive control (MPC) for high degree of freedom (DoF) robots. 
Experimental results demonstrate that by parameterizing the input trajectory more than three quarters of the optimization variables used in traditional MPC can be eliminated with practically no effect on system performance. 
This parameterization also leads to trajectories which are more conservative, producing less overshoot in underdamped systems with modeling error. In this paper we present two MPC solution methods that make use of this parameterization. The first uses a convex solver, and the second makes use of parallel computing on a graphics processing unit (GPU). 
We show that both approaches drastically reduce solve times for large DoF, long horizon MPC problems allowing solutions at real-time rates.
Through simulation and hardware experiments, we show that the parameterized convex solver MPC has faster solve times than traditional MPC for high DoF cases while still achieving similar performance. For the GPU-based MPC solution method, we use an evolutionary algorithm and that we call Evolutionary MPC (EMPC). EMPC is shown to have even faster solve times for high DoF systems. Solve times for EMPC are shown to decrease even further through the use of a more powerful GPU. This suggests that parallelized MPC methods will become even more advantageous with the improvement and prevalence of GPU technology.
%Finally, simulation experiments demonstrate that parameterizing and parallelizing the MPC problem enables a real-time global search of the nonlinear MPC problem.
%Hardware experiments also demonstrate that EMPC solve times are less affected by estimation and message passing processes which compete with conventional MPC for CPU time.

\end{abstract}

\begin{IEEEkeywords}
Optimal Control, Model Predictive Control, Graphics Processing Unit, Convex Optimization, Genetic Algorithms
\end{IEEEkeywords}

\section{Introduction}
As applications for robotics and automation increase, so will the need for robot platforms which are able to perform well in not just one, but  a variety of tasks. This versatility usually comes at the cost of simplicity - leading to more versatile, yet complex robots with many Degrees of Freedom (DoF). Often the design of these complex robots mimics the design of biological systems such as humanoid robots \cite{collins2001three}, \cite{hubicki2016atrias}. These high DoF robots are technically capable of performing the complex dynamic behaviors such as those we see in nature, however their control is very difficult due to the many system states and inputs which are coupled in nonlinear ways.

Optimal control ideas have proven useful in modeling high DoF biological systems and in reproducing life-like motion with robots \cite{Li2004}, however the computational complexity of many optimal control algorithms often prohibits their real-time use on systems with many DoF. Because of this, there is a significant amount of active research in how to solve optimal control problems associated with high DoF control quickly enough to perform real-time control.

In this work we propose a method of control trajectory parameterization which can be used to extend the tractability of Model Predictive Control (MPC - a form of optimal control) to systems with more DoF or to lengthen the look-ahead horizon of MPC. We demonstrate that the proposed method drastically decreases MPC solve times while having little effect on performance and potentially improving robustness. We detail and compare two methods used for solving the modified MPC problem. We call these methods parameterized convex solver MPC and evolutionary MPC (EMPC). While parameterized convex solver MPC uses a fast convex solver like many implementations of traditional MPC, EMPC uses the parallel computation available in graphics processing units (GPUs) in order to solve the optimization.

Specifically the contributions of this paper are:

\begin{itemize}
    \item Presentation of experimental results demonstrating the effects of control trajectory parameterization on MPC performance and robustness for robot control.
    \item The development of a parameterized version of MPC which can be solved using a fast convex solver
	\item The development of a very flexible parallelized and parameterized version of MPC based on an evolutionary optimization algorithm - Evolutionary MPC (EMPC).
	\item Comparisons of EMPC, parameterized convex MPC, and traditional MPC both in simulation and on real robot hardware.
\end{itemize}

%Model Predictive Control (MPC) is an optimal control algorithm which was originally developed for the chemical processing industry, but thanks to advancements in computers and fast optimization solvers, has become increasingly popular in recent years in robotics applications. It allows for tasks to be defined in terms of intuitive cost functions and constraints on states and inputs.

%As computers become more powerful, more complex control algorithms are made possible. MPC used to be only realistic for systems with slow dynamics, but is now becoming more and more popular for robotics. 
%While LQR has an analytical solution which allows it to be solved very quickly, it is limited to linear (or linear time varying) systems and cannot take into account constraints in state or input. MPC allows for control very similar to LQR in the absence of constraints, but allows for definition of constraints as well as more flexibility in terms of defining cost and dynamics. 
%MPC in its most general form can be used to find optimal controls for systems with arbitrarily complex cost, dynamics and constraints, however due to the need for real-time control the models and cost functions are typically kept pretty simple for MPC. People have addressed these problems with coupling torque methods to break up the dynamic system into several smaller systems and by improving the efficiency of the optimization at the heart of MPC. Methods for making MPC more tractable at long horizons and more complex systems is still an active area of research.

The remainder of this paper is organized as follows: Section \ref{related_work} highlights related work done in the fields of high DoF control and MPC parameterization and parallelization. Section \ref{parameterized_mpc} explains how control trajectories are parameterized for use with MPC, as well as the effects of this parameterization on MPC performance and robustness. Section \ref{parameterized_convex_mpc} explains two implementations of parameterized MPC which can be solved using a convex solver as well as experiments designed to test the solve times of each. Section \ref{empc} explains the implementation of a parameterized and GPU parallelized variant of MPC (EMPC). Section \ref{comparing_mpc_methods} contains simulation and hardware experiments which highlight the advantages and drawbacks of the proposed methods as well as comparisons to traditional MPC methods. Section \ref{conclusion} summarizes our findings and proposes future work.

\section{Related Work}\label{related_work}

\subsection{Optimal Control of High DoF Systems}

A well studied and intuitive method of performing optimal control for a high DoF robot is to take a hierarchical approach. Using the Operational Space Formulation \cite{Khatib1987},
one may prioritize tasks to be completed and ensure that lower priority tasks are only executed in the null space of higher priority task control. In \cite{Sentis2006} as well as \cite{Dietrich2012} this is used to control a humanoid robot, while in \cite{Hutter2014} it is used as part of a locomotion scheme for a quadruped. In \cite{Escande2014} a similar control hierarchy is achieved through quadratic programming which also allows for control of a humanoid.

Another way to perform optimal control in a hierarchical way is the approach taken in \cite{Kuindersma2016} where a high level optimization is done to find footsteps for the humanoid robot Atlas, a lower level optimization is done to plan joint trajectories given simplified dynamics \cite{Dai2014}, and an even lower level LQR controller is used to track those joint trajectories. This stacking of optimizations allows simplifications which make each layer tractable while still finding optimal or near-optimal solutions for the full high DoF problem.

Alternatively, methods such as iLQR/DDP \cite{Li2004} \cite{Tassa2014} work by forward simulating nonlinear dynamics given an input trajectory, and then using derivative information about the cost and the dynamics to calculate an improvement to the trajectory. These methods have been shown to work on high DoF systems with nonlinear dynamics (such as humanoid robots) \cite{Koenemann2015}, however there is still a high computational cost associated with forward simulating nonlinear dynamics and calculating its derivatives. The need to calculate these dynamics and derivatives quickly has even driven the development of software specially designed to do these at speeds which allow for MPC (MUJOCO) \cite{Todorov2012}.

The literature on walking robots is rich with examples of high DoF controllers, many based on optimal control ideas. In \cite{Pratt2001} Virtual Model Control is presented as a method of simplifying the high DoF system into a lower DoF system through feedback. The lower DoF system can then be controlled using standard optimal control techniques. Building on the Zero Moment Point (ZMP) preview control approach to walking \cite{Kajita2004}, several approaches use MPC as a method to control the ZMP toward some desired trajectory or point \cite{Krause2012}, \cite{Wieber2009}. In the Springer Handbook of Robotics \cite{springer_handbook} they claim that the motion generation schemes which power most of the great humanoids use MPC in one form or another.

Advances in direct optimization methods for MPC have come from advances in convex optimization solvers such as OSQP \cite{osqp} and CVXGEN \cite{cvxgen}, as well as researchers exploiting known structures in optimal control problems \cite{Kuindersma2014efficient}. Fast QP solvers have grown to handle larger and larger problems and have also decreased the time taken to find solutions, allowing us to perform MPC for larger and more complex systems at higher rates or with longer horizons. Fast direct solvers usually assume linear dynamics, however there are solvers which admit nonlinear dynamics at the cost of more computation time. There has even been research done in order to allow direct solvers to handle contact dynamics \cite{Posa2013}.

There are also methods to decrease the problem size and complexity to allow convex solvers not only to solve at real-time speeds, but to avoid infeasible problems which will fail to solve at all \cite{Rossiter2010}. In \cite{jon_terry} a modeling method is developed which decouples portions of the system, allowing each portion to be controlled independently using separate MPC controllers. In \cite{Ling2012} and \cite{Ling2011} a similar idea is used, assuming that each input acts independently and therefore can be optimized separately. These controllers exhibit better disturbance rejection because they do something good soon, rather than something better later.

%Other optimal control methods used for walking include optimally tracking a trajectory given by biological examples \cite{ilqr_gps_paper} and optimally tracking trajectories given by Poincare maps \cite{optimal_poincare_walking}. Poincare maps themselves are often found by minimizing some cost criteria such as energy \cite{passive_walking_maybe?}.

Some high DoF robots have been successfully controlled using different forms of Reinforcement Learning in order to mimic the behavior of optimal controllers. Guided policy search \cite{Levine2013} for example, trains a Neural Network using examples from a nonlinear trajectory optimization (DDP). Because the training does not need to happen in real-time, this allows guided policy search to learn to approximate a nonlinear optimal controller. Although the optimal controller may not be able to solve at real-time rates, the execution times for machine learned models are typically very fast, allowing for very fast nonlinear control. This method has been used for humanoid walking, complex contact-rich tasks \cite{Levine2015contact}, and even learning visuomotor policies \cite{Levine2015}, which are all high DoF tasks.

The pattern that we see in the literature is that in order to solve the MPC problem at fast enough rates for real-time control, some simplifying assumptions or approximations must be made. In this work we present a novel assumption which allows for MPC solutions at higher rates and with longer horizons. The unique assumption made in this work is that a time-varying control trajectory may be represented as a linear function of relatively few equally spaced points. We believe that this novel approach may be combined with others in the literature in order to decrease solve times or extend MPC to more high DoF systems. We present two methods in this work which use this assumption to solve the MPC problem at high rates. The effects of this assumption on MPC performance and robustness are discussed in Sections \ref{parameterized_mpc_performance_results} and \ref{parameterized_mpc_robustness_results}.

\subsection{Parallelized and Parameterized MPC}

Recently, parameterization methods have begun to gain attention a way to reduce the complexity of MPC. In \cite{Muehlebach2019} both inputs and states are parameterized, taking advantage of known properties. In \cite{Khan2013} orthogonal basis polynomials are explored as a form of parameterization for MPC, while in \cite{Lengagne2013} B splines are used to represent states and inputs in an MPC optimization.

Dynamic Movement Primitives (DMPs) represent motions or behaviors using stable nonlinear attractor functions \cite{schaal2006dynamic} instead of time-based trajectories. This idea has been used to model the complex motions seen in humans \cite{kulic2012incremental} and to compile libraries of ``skills" which can be used to generate movements in robots \cite{konidaris2012robot}. A major advantage of DMPs is that they represent closed loop behaviors which are more robust to disturbances. This makes them well suited to real world tasks in unstructured environments such as furniture assembly \cite{niekum2015learning}.

In \cite{mukadam2018continuous} continuous time trajectories are modeled as Gaussian processes. This allows the entire trajectories to be represented using a small number of states and enables very fast interpolation needed for fast planning. This is similar to the idea of parameterization found in \cite{paraschos2013probabilistic} where the idea of a probabilistic motion primitive is introduced.

One popular MPC method involves solving the iLQR trajectory optimization problem rapidly and using the input and/or feedback policy from the first time step in an MPC scheme. Recently, work has been done to parallelize the computations needed for iLQR. iLQR can be thought of as a single-shooting method for solving the initial value problem defined by the dynamics constraint, followed by a backwards policy update using Ricatti recursions. In \cite{Giftthaler2017} the authors use a multiple-shooting method to forward simulate sections of the dynamic trajectory in parallel, followed by a backwards policy and state trajectory update. This allows for faster solution times and is more robust to poor initial guesses. This has been shown to successfully control a quadruped \cite{Neunert2017}.

In very early work in parallelized MPC, the authors of \cite{Rogers2013} and \cite{Ilg2011} used sampling-based optimization methods on a GPU to find trajectories for parafoils and projectiles which were robust to wind disturbances. More recently, the authors of \cite{Williams2016} and \cite{Williams2017} used a parallelized policy improvement method seeded with a learned policy to run MPC on a GPU at real-time rates. Our EMPC method is most similar to this method, however our method of parameterizing the input space makes EMPC tractable without the need for a prior policy to improve upon. The reduction in the search space afforded by the control trajectory parameterization means that a global search is possible using a heuristic global optimization method (the evolutionary algorithm). This is in contrast to gradient-based or policy improvement methods, which start at an initial point in the search space and descend to a local minima.

\section{Effects of Control Trajectory Parameterization on MPC}
\label{parameterized_mpc}

\subsection{Brief Review of Model Predictive Control}
Given a linear or linearized system, we can describe the system dynamics in state variable form as

\begin{equation}
    \mathbf{\dot{x} = Ax + Bu + w}
\end{equation}

where $\mathbf{x}$ is the vector of states, $\mathbf{u}$ is the vector of system inputs, and $\mathbf{w}$ is a vector of constant disturbances. If there are $n$ states and $m$ inputs, then the matrices $\mathbf{A}$ and $\mathbf{B}$ are $n$x$n$ and $n$x$m$ respectively, while $\mathbf{w}$ is an $n$x1 vector.

Using any discretization method (Euler, semi-implicit Euler, matrix exponential, etc.) we can create a discretized state space model:

\begin{equation}
\label{eq:first_order_integration}
    \mathbf{x_{k+1} = A_dx_k + B_du_k + w_d}.
\end{equation}

The above equation can be used to forward simulate the states of our system, given initial conditions and inputs. In traditional MPC these discretized dynamic equations will become the constraints of our optimization. In an MPC solver looking forward over a horizon of $T$ time steps, the optimization may be formulated as:

\begin{equation}
\label{eq:mpc_problem}
\begin{split}
    J &= \sum_{k=0}^{T-1} \bigg[(\mathbf{x_{goal}-x_{k})}^T Q (\mathbf{x_{goal}-x_{k})} \\
    &+ (\mathbf{u_{goal}-u_k)}^T R (\mathbf{u_{goal}-u_k)}\bigg]\\
    &s.t. \\
    &\mathbf{x_{k+1} = A_dx_k + B_du_k + w_d}\quad \forall \quad k=0,...,T\\
    &\mathbf{x_{min} \leq x_k \leq x_{max}} \quad \forall \quad k=0,...,T \\
    &\mathbf{u_{min} \leq u_k \leq u_{max}} \quad \forall \quad k=0,...,T \\
\end{split}
\end{equation}

where $J$ is the objective function value, $\mathbf{x_{goal}}$ and $\mathbf{u_{goal}}$ are the goal states and inputs respectively, and  $\mathbf{x_{min}}$, $\mathbf{x_{max}}$, $\mathbf{u_{min}}$, $\mathbf{u_{max}}$, are the state and input bounds. For all of the experiments in this work $\mathbf{u_{goal}}$ is defined as the zero vector, meaning the cost is quadratic on the input and weighted by the matrix $R$.

MPC solves the above optimization for the entire horizon of length $T$, however only the first input ($\mathbf{u_0}$) is applied to the system. After applying this input, the optimization is solved again using new state and model information. This process is repeated with MPC only ever applying the first input, but solving over an entire horizon of value $T$.
%While the entire open loop sequence of $\mathbf{u_k}$s could be applied, this would lead to poor performance in the case of modeling error or un-modeled disturbances.

Measures of control performance such as rise time, settling time, and percent overshoot can be affected by altering the weighting matrices $Q$ and $R$, as well as the horizon length $T$. These quantitative performance measures however are not explicitly incorporated into the objective function and therefore solutions which are technically optimal may have higher rise times, settling times, and overshoot. In order to make fair comparisons between MPC solutions, when evaluating the performance of MPC for the rest of this paper, we choose to evaluate the same objective function used when solving the MPC problem, but instead using the actual inputs and resulting states on the real system over a defined amount of time. The actual cost evaluated over $H$ time steps may be stated as 

\begin{equation}
    \label{eq:actual_cost}
    Actual\ Cost = \sum_{i=0}^{H} J(\mathbf{x_{i}},\mathbf{u_{i}})
\end{equation}
where $\mathbf{x_{i}}$ and $\mathbf{u_{i}}$ represent the values of the state and input at time step $i$ and $J(\cdot, \cdot)$ represents the objective function. Throughout the rest of this paper we will refer to the cost calculated using actual states and inputs applied to the system as ``actual cost.''

\subsection{Method of MPC Parameterization}
\label{mpc_parameterization_method}
Traditionally, the optimization defined in Equation \ref{eq:mpc_problem} is solved by finding separate values of $\mathbf{u_k}$ for each discrete time step $k$. This amounts to an optimization over $T*m$ variables, or an optimization in a $T*m$ dimensional space. This optimization becomes difficult to solve fast enough for real-time control over long horizons (large $T$) or for systems with many inputs (large $m$). Reducing the dimension of this search space is the goal of control trajectory parameterization.

By parameterization of a control trajectory, we mean a method to represent the inputs of a time varying control trajectory using fewer than $T*m$ parameters. This is similar to the idea of curve fitting, where many discrete data points are represented by a smaller set of numbers such as coefficients of a polynomial expression, or the Fourier Transform which represents a signal using a small set of coefficients of trigonometric functions.

\begin{figure}
    \centering
    \includegraphics[width=\linewidth]{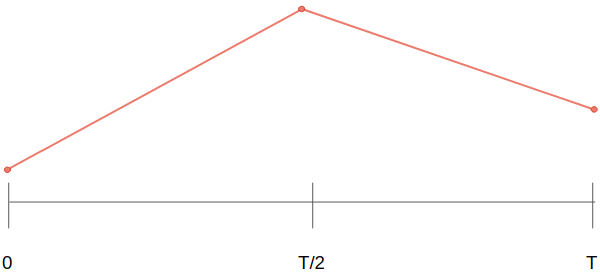}
    \caption{An example of a control trajectory over T time steps parameterized by three ``knot" points.}
    \label{fig:one_parameterized_line}
\end{figure}

Instead of choosing to represent control trajectories using coefficients of polynomial or trigonometric functions, we choose to represent them using piece wise linear functions which cross through equally spaced ``knot" points as seen in Figure \ref{fig:one_parameterized_line}. This parameterization is fairly intuitive, allows for simple bounding of the control trajectory, and also preserves the convexity of the optimization problem. We do not intend to claim that this method of parameterizing MPC is the best form of parameterizing the input trajectory. However, experimental results in Section \ref{parameterized_mpc_performance_results} demonstrate that performance comparable to un-parameterized MPC can be achieved by using a sufficient number of knot points.

In the case of a convex cost function with affine constraints, it is possible to show that there exists one optimum which is both the local and global optimum \cite{boyd2004convex}. This means that linear MPC, when implemented with a perfect model on a linear system with a sufficiently long horizon, will execute the optimal control at each time step and will achieve the optimal actual cost. While it is possible for parameterized MPC to also achieve the optimal actual cost, it can never achieve a lower actual cost. Again, this is only true for linear MPC executed on a linear system with a sufficiently long horizon.

\subsection{Parameterized MPC Performance - Experimental Setup}
\label{parameterized_mpc_performance_setup}

In order to evaluate the performance of MPC using a parameterized control trajectory, we conduct experiments on three simulated systems of varying complexity: a pendulum unaffected by gravity (linear), a pendulum in the presence of gravity (nonlinear), and a Puma 560 robot (nonlinear). 

\subsubsection{Pendulum}
The continuous time dynamics of an inverted pendulum are
\begin{equation}\label{eq:inverted_pendulum}
    m l^2 \ddot{q} + b \dot{q} + m g l sin(q)  = \tau_{motor} 
\end{equation}
where $m$ is the mass at the end of the link, $l$ is the length of a massless link, $b$ is a viscous damping coefficient, and $g$ is the acceleration of gravity. For experiments without gravity we let $g=0$. 
%We choose as our state $x = [\dot{q}; q]$.

\subsubsection{Puma 560}
We assume the robot is comprised of rigid links and pin joints, so that the dynamic equations take the canonical form 
\begin{equation}\label{eq:puma_dynamics}
    M(q)\Ddot{q} + C(q,\dot{q}) + b\dot{q} + \tau_{grav} = \tau
\end{equation}
where $q$ is the vector of generalized coordinates, $M(q)$ is a configuration dependent inertia matrix, $C(q,\dot{q})$ represents torques produced by centrifugal and Coriolis forces, $b$ is a viscous damping coefficient, $\tau_{grav}$ are the torques applied by gravity on the robot and $\tau$ are applied torques from the motors. 
%We choose as our state $x = [\dot{q}; q]$.
%\todo{Do we need to explain our linearization? I'd rather not personally...}

In order to isolate the effect of parameterization on the MPC problem, initial experiments were performed in simulation using MATLAB's fmincon function to perform the optimization for MPC. MPC is used to control the simulated system for one second from an initial position at rest, to a goal position at rest. To avoid any bias from a particular part of the robot's workspace, we ran several trials with initial and goal positions sampled from a random uniform distribution and we report statistics over all trials.

The metric that we use to evalute the effect of parameterization on MPC performance is the actual cost calculated by evaluating the cost function over the one second simulation using actual states and inputs. However because trials are run for different initial and different goal states, we would expect a large amount of variation in costs even using the same MPC controller. To eliminate this variation due to different initial and goal positions, we normalize the actual cost of each trial by dividing it by the actual cost achieved by traditional MPC with a horizon equal to the actual simulation time (one second). We would expect this ratio to be greater than one for most cases because traditional MPC with a long horizon should find an optimal or near-optimal solution compared to our parameterized version. A ratio of less than one indicates that the performance is better than traditional MPC with a long horizon. For reference, the actual cost ratio described above, evaluated over $H$ time steps can be expressed as

\begin{equation}
    \label{eq:actual_cost_ratio}
    Actual\ Cost\ Ratio = \frac{\sum_{i=0}^{H} J(\mathbf{x_{param,i}},\mathbf{u_{param,i}})}{\sum_{i=0}^{H} J(\mathbf{x_{trad,i}},\mathbf{u_{trad,i}})}
\end{equation}

where $\mathbf{x_{param,i}}$ and $\mathbf{u_{param,i}}$ represent the values of the state and input at time step $i$ using a parameterized MPC, while $\mathbf{x_{trad,i}}$ and $\mathbf{u_{trad,i}}$ represent the values of the state and input at time step $i$ using traditional MPC.

For each of the robots, we use the setup described above and vary the number of parameters (knot points) used in parameterized MPC with a set horizon (50 time steps) in order to control the robot from the initial position to the goal position. We simulate running MPC at a rate of 100 Hz with a model that has been discretized at a time step of .01 s.

Using our piecewise linear parameterization we can use anywhere between one and $Tm$ parameters where $m$ is the number of inputs. In fact if we use $Tm$ parameters, then the problem and solution are identical to those in traditional MPC. When using $Tm$ parameters, the distance between points is exactly the same as the discretization time step and each knot point becomes the input applied over the discrete time interval, just as in traditional MPC.
%the last knot point (corresponding to $m$ parameters) is never applied to the system. 
By varying the number of parameters between one and $Tm$, we are able to clearly see the effect that using fewer parameters has on MPC performance. These results are shown in Figure \ref{fig:num_parameters_vs_log_cost}.

For comparison, we also ran experiments varying the horizon length for traditional MPC. This provides a useful context for reasoning about performance because horizon length is commonly used as a tuning parameter. Horizon length can also be shortened until the MPC problem can be solved fast enough for real-time control, often at the expense of performance. The results of the experiments varying horizon length are found in Figure \ref{fig:horizon_vs_log_cost}. 

\subsection{Parameterized MPC Performance - Experimental Results}
\label{parameterized_mpc_performance_results}
\begin{figure}
    \centering
    \includegraphics[width=1\linewidth]{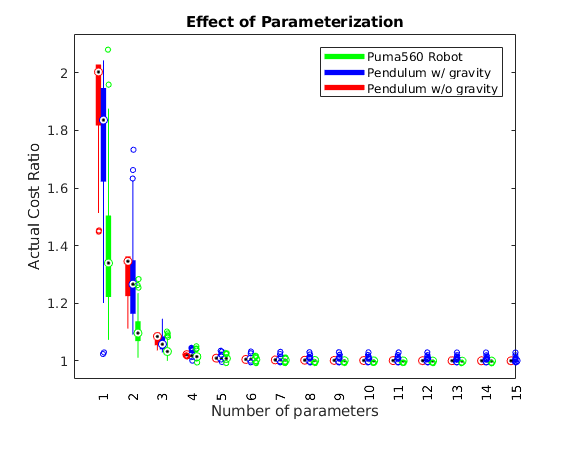}
    \caption{Ratio of actual cost using parameterized MPC with a given number of parameters to traditional MPC for different robot platforms.}
    \label{fig:num_parameters_vs_log_cost}
\end{figure}

The results of the first experiment can be seen in the box plot in Figure \ref{fig:num_parameters_vs_log_cost}. All box plots in this work follow the following convention: the box contains data between the first and third quartile, the median is represented by a marker within the box, and whiskers contain all data not considered outliers (more than 1.5 times the interquartile range away from the box). 

The effect on cost of decreasing the number of parameters is very slight until the number of parameters reaches a threshold between three and four points. This suggests that the important features of the optimal control trajectory are able to be represented fairly well with four or more points using our simple parameterization. The cost is of course the highest for the parameterization with one point. This corresponds to the optimization picking one input which will be applied over the entire horizon.

It can be seen that the actual cost ratio never goes below one for the linear pendulum case. This is the expected result because the long horizon traditional MPC has a perfect model which is valid over the entire horizon and so has found the actual optimal control solution. As we increase the number of parameters it can be seen that parameterized MPC quickly converges to the same solution. 

The median cost ratio for the nonlinear pendulum is seen to be lower than that of the linear pendulum for the lower order parameterizations and the cost ratio is also seen to dip below one occasionally. 
%This indicates that there are times when parameterized MPC performs slightly better than traditional MPC. 
Traditional linear MPC is not expected to find the absolute optimal solution for a nonlinear system because the linearized model that is used for optimization is only valid for a small region surrounding the linearization point. The fact that the cost ratio is generally lower for the nonlinear pendulum than the linear pendulum does not necessarily indicate that parameterized MPC does better with nonlinear systems, but rather that it is more comparable to traditional MPC for nonlinear systems. This is likely because traditional linear MPC does not perform as well for nonlinear systems as it does for linear systems. This effect is further demonstrated with the results on the Puma 560 robot, which is an even more nonlinear system.
%The fact that parameterized MPC occasionally outperforms traditional MPC may be dependent on the specific part of the workspace or some other circumstance. This will be discussed in greater length in Section \ref{parameterized_mpc_robustness}. 

\begin{figure}
    \centering
    \includegraphics[width=\linewidth]{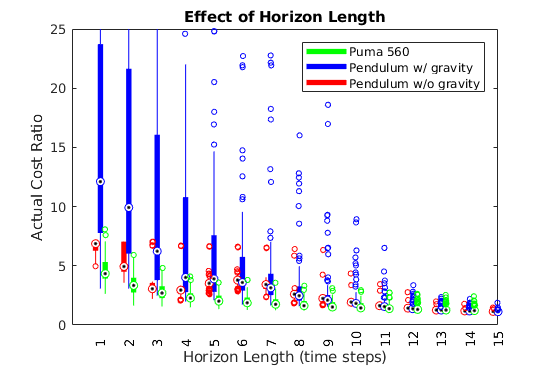}
    \caption{Ratio of actual cost using traditional MPC with a given horizon to traditional MPC with a horizon of 50 for different robot platforms.}
    \label{fig:horizon_vs_log_cost}
\end{figure}

The results found by varying the horizon length as seen in the box plot in Figure \ref{fig:horizon_vs_log_cost} are not surprising. As expected, a shorter horizon length encourages more greedy behavior, which leads to a less optimal solution. 
%There is also a clear trend which indicates that increasing horizon length past a certain threshold does not lead to large performance gains. 
The purpose of this experiment however, is to provide a context when reasoning about MPC performance. When we compare the scale of Figures \ref{fig:num_parameters_vs_log_cost} and \ref{fig:horizon_vs_log_cost} we observe that the effect of decreasing horizon is far more dramatic than that of decreasing the number of parameter knot points.
%\todo{can we quantify this?}

This is an important result which should influence the design of MPC controllers. It indicates that instead of shortening the horizon length of traditional MPC until solve times allow for real-time control, we should often instead parameterize the control space to reduce solve times. The effect of parameterizing the control space is often smaller than the effect of decreasing horizon length, especially if the horizon length is not long to start with. %These effects, while true for the linear pendulum, seem even more pronounced with nonlinear systems.
%This comparison between the effects of parameterization and horizon length also validates our first hypothesis: ``The difference in solution quality between traditional MPC and MPC using a parameterized control trajectory is negligible." 
%While the effect of using any more than five parameters is truly negligible, when compared to the effects of shortening horizon length, even the effect of using a single input over the horizon is generally smaller.

%Note that the results presented in this section are all found using a piecewise linear parameterization. While improvements could be made by using alternate parameterizations, we note that the performance of a better parameterization would necessarily lie between those of our parameterization and traditional MPC. Because the results show that this gap in performance is very small for a sufficient number of parameters, any gain in performance from using another parameterization must be even smaller.

%\subsection{Effects of Control Trajectory Parameterization on MPC Robustness}

\subsection{Parameterized MPC Robustness - Experimental Setup}\label{parameterized_mpc_robustness_setup}
While specific MPC methods have been developed aimed at improving robustness \cite{bemporad1999robust} \cite{ostafew2016robust}, traditional MPC itself has been shown to be robust to modeling error and disturbances. The experiments carried out in this section seek to determine the effect of parameterization on the inherent robustness of MPC.

In order to experimentally test the robustness of our controllers to modeling error we intentionally introduce error into the model used for MPC. We then use MPC with an incorrect model to control the simulated system for which we know the model perfectly. Error is introduced in the form of an ``error multiplier" which is multiplied by certain model parameters. When the error multiplier is equal to one there is no modeling error, while values less than or greater than one correspond to underestimates or overestimates of model parameters respectively. For experiments using the inverted pendulum, the error multiplier was applied to both the mass and length of the pendulum. For experiments using the Puma robot, the error multiplier was applied to the entire inertia matrix.

We first quantify the effect of modeling error on MPC performance using the ``actual cost" metric defined in Equation \ref{eq:actual_cost}. In order to compare the sensitivity of several controllers to modeling error, we normalize the actual cost incurred by each controller by dividing by the actual cost incurred by a controller without modeling error. The normalized cost can be stated as

\begin{equation}
    \label{eq:actual_cost_ratio}
    Normalized\ Cost = \frac{Actual\ cost_{with\ error}}
    {Actual\ cost_{without\ error}}.
\end{equation}

This metric shows the sensitivity of MPC performance to modeling error, however we are also interested in finding the sensitivity of MPC \textit{stability} to modeling error. In general, it is difficult to prove the stability of an MPC controller without the use of a local stabilizing controller and terminal constraints as well as costs \cite{mayne2000constrained}, \cite{friedbaum2018model}. These proofs also often operate under the assumption of accurate modeling, which is not the case in these experiments. It is generally the case however that conservative controllers tend to be more stable than aggressive controllers. We choose to quantify the ``conservative-ness" of a controller by measuring the rise time and percent overshoot attained using that controller. Aggressive controllers produce small rise times and large percent overshoot, while conservative controllers produce larger rise times and smaller percent overshoot.

\subsection{Parameterized MPC Robustness - Experimental Results}\label{parameterized_mpc_robustness_results}

\begin{figure}
    \centering
    \includegraphics[width=\linewidth]{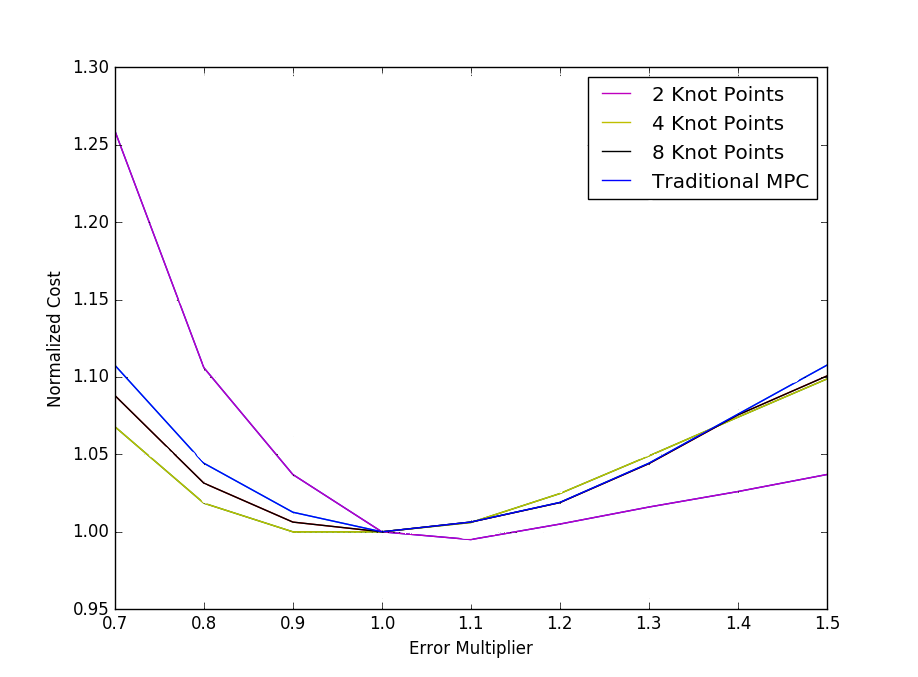}
    \caption{How traditional MPC compares to parameterized MPC when an error multiplier is applied for an inverted pendulum. Due to exponential increase with error multiplier values of .5 and .6, these values are omitted from this figure to emphasize how a system is affected closer to an error multiplier of 1 (+ or - 10-30\% modeling error for a system).}
    \label{fig:normalized_cost_inverted_pendulum_param_v_vanilla}
\end{figure}

\begin{figure}
    \centering
    \includegraphics[width=\linewidth]{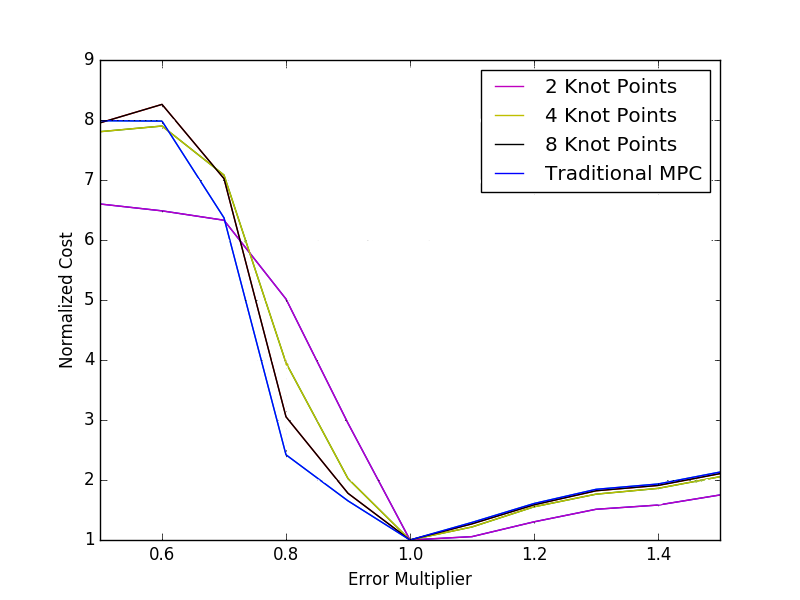}
    \caption{How traditional MPC compares to parameterized MPC when an error multiplier is applied for the Puma 560 robot.}
    \label{fig:normalized_cost_puma560}
\end{figure}

Comparisons of parameterized MPC to traditional MPC performance sensitivity can be seen in Figures \ref{fig:normalized_cost_inverted_pendulum_param_v_vanilla} and \ref{fig:normalized_cost_puma560}. The results from Figure \ref{fig:normalized_cost_inverted_pendulum_param_v_vanilla} reveal that the sensitivity of MPC performance using parameterized MPC with more than two parameters is similar to that of traditional MPC. While parameterized MPC performance using two knot points is less sensitive to overestimating inertial parameters, it is more sensitive to underestimating them. As more knot points are used in parameterized MPC, the sensitivity approaches that of traditional MPC. This is the expected result because the parameterized MPC is more closely approximating traditional MPC. We do not believe that these results conclusively show that either parameterized or traditional MPC performance is less sensitive to modeling error than the other. In other words, parameterizing MPC does not increase or decrease the sensitivity of MPC performance to modeling error.

Figure \ref{fig:normalized_cost_puma560} serves to confirm that the results found in Figure \ref{fig:normalized_cost_inverted_pendulum_param_v_vanilla} hold for a more complex system (the Puma robot). It also exposes an interesting trend in MPC sensitivity which is true for both parameterized and traditional MPC. While MPC performance degrades very quickly for the case of underestimated inertia, it degrades relatively slowly in the case of overestimated inertia. This information should inform the modeling of dynamic systems for use with MPC. Specifically, mass and length estimates should err on the side of overestimating inertia when used for MPC because underestimating inertia has a strong negative impact on MPC performance.

The results of quantifying percent overshoot and rise time in the inverted pendulum experiments can be seen in Tables \ref{tab:percent_overshoot} and \ref{tab:rise_time}. In both tables it can be seen that when inertia is overestimated (error multiplier greater than one), the MPC controllers act more conservatively and reduce or eliminate overshoot while increasing rise times. Underestimating the inertia (error multiplier less than one) has the opposite effect, leading to greater overshoot and shorter rise times for the case of traditional MPC and parameterized MPC with 8 knot points. This result agrees with the results seen in Figures \ref{fig:normalized_cost_puma560} and \ref{fig:normalized_cost_inverted_pendulum_param_v_vanilla}. Interestingly, parameterized MPC using two or four knot points does not strictly decrease rise time when underestimating inertia.

There is a clear trend which indicates that by parameterizing with fewer knot points, rise time is increased and percent overshoot is decreased. This trend seems consistent for all modeling errors included in these experiments. In other words, using fewer knot points in parameterized MPC results in more conservative control. This result makes sense intuitively because within a prediction horizon, a parameterized MPC controller cannot change inputs instantaneously. Within the prediction horizon inputs must change linearly and can only change direction at knot points. This makes very aggressive ``bang-bang" maneuvers impossible within the MPC prediction horizon. Thus by parameterizing the control trajectory of MPC we have enforced a certain amount of ``conservative-ness".

Because conservative controllers are generally more stable than aggressive controllers when modeling error or disturbances are introduced, the inherent conservative nature of parameterized MPC favors stability over aggressiveness. This also helps to explain the results found in Section \ref{parameterized_mpc_performance_results} where parameterized MPC was found to perform slightly worse than traditional MPC in most cases, especially with few knot points. The decrease in performance can at least partially be attributed to the fact that the parameterized MPC controller was more conservative and therefore incurred higher cost.

As a partial summary, in Section \ref{parameterized_mpc_performance_results} it is shown that the performance difference between parameterized and traditional MPC is relatively small for a sufficient number of knot points. Results from Section \ref{parameterized_mpc_robustness_results} demonstrate that this difference in performance is at least partially because parameterized MPC favors more conservative control. These results together suggest that the performance of parameterized MPC can be very close to that of traditional MPC and, especially when a model is not known very well, will favor more conservative control actions leading to more stable control.

\begin{table}
\begin{center}
 \begin{tabular}{|m{3em}|| m{1cm} | m{1cm} | m{1cm} | m{1cm}|} 
 \hline
 \multicolumn{5}{|c|}{Percent Overshoot for Parameterized Systems} \\
 \hline
 Error Multiplier & 2 Knot Points & 4 Knot Points & 8 Knot Points & Traditi- onal MPC \\ [0.5ex] 
 \hline\hline
 .5 & 37.7 & 40.2 & 40.3 & 43.5\\ 
 \hline
 .6 & 26.5 & 29.3 & 29.3 & 31.1 \\
 \hline
 .7 & 18.6 & 19.8 & 20.8 & 22 \\
 \hline
 .8 & 12.1 & 12.3 & 13.8 & 14.6 \\
 \hline
 .9 & 6.88 & 6.78 & 8.13 & 8.64 \\ 
 \hline
 1 & 3.05 & 2.96 & 3.89 & 4.17 \\
 \hline
 1.1 & .772 & .732 & 1.18 & 1.3 \\
 \hline
 1.2 & .0252 & .0171 & .0729 & .0919 \\
 \hline
 1.3 & 0 & 0 & 0 & 0 \\
 \hline
 1.4 & 0 & 0 & 0 & 0 \\
 \hline
 1.5 & 0 & 0 & 0 & 0 \\[1ex] 
 \hline
\end{tabular}
\caption{\label{tab:percent_overshoot}Data on the percent overshoot for an inverted pendulum, comparing traditional MPC simulation to different parameterized MPC simulations.}
\end{center}
\end{table}

\begin{table}
\begin{center}
  \begin{tabular}{|m{3em}|| m{1cm} | m{1cm} | m{1cm} | m{1cm}|} 
 \hline
 \multicolumn{5}{|c|}{Rise Time for Parameterized Systems} \\
 \hline
 Error Multiplier & 2 Knot Points & 4 Knot Points & 8 Knot Points & Traditi- onal MPC \\ [0.5ex]  
 \hline\hline
 .5 & .85 & .41 & .3 & .28 \\ 
 \hline
 .6 & .65 & .36 & .3 & .29 \\
 \hline
 .7 & .56 & .35 & .31 & .3 \\
 \hline
 .8 & .54 & .36 & .32 & .31 \\
 \hline
 .9 & .54 & .38 & .35 & .34 \\
 \hline
 1 & .59 & .43 & .39 & .38 \\
 \hline
 1.1 & .67 & .5 & .45 & .44 \\
 \hline
 1.2 & .78 & .6 & .55 & .54 \\
 \hline
 1.3 & .89 & .71 & .67 & .66 \\
 \hline
 1.4 & .96 & .81 & .77 & .76 \\
 \hline
 1.5 & 1.02 & .88 & .85 & .84 \\[1ex] 
 \hline
\end{tabular}
\caption{\label{tab:rise_time}Data on the rise time for an inverted pendulum, comparing traditional MPC simulation to different parameterized MPC simulations.}
\end{center}
\end{table}

\section{Parameterized Convex Solver MPC}
\label{parameterized_convex_mpc}

While MPC has been used in many domains such as the chemical process industry \cite{binder2001introduction}, the focus of this work is on the application of MPC to robotics. In order to use MPC in real-time for robots, it is necessary to solve the optimization problem very quickly. For this reason, fast convex solvers have become the method of choice for use with MPC in robotics. 

In this section we detail a method for implementing a parameterized version of MPC for use with a convex solver. In addition to outlining the method of implementation, we also provide experimental results demonstrating the effects of this parameterization on MPC solve times. Because the parameterization is the same as in Section \ref{mpc_parameterization_method}, the effects on performance and robustness remain the same as in Sections \ref{parameterized_mpc_performance_results} and \ref{parameterized_mpc_robustness_results}.

Formulating the MPC problem to fit within a convex solver framework is somewhat restrictive, since the optimization must be convex with only linear equality or inequality constraints. However, there are still a couple of ways to formulate the problem. Most convex solvers admit problems of the form

\begin{equation}\label{eq:convex_solver_form}
\begin{split}
    minimize& \ \ \  z^T P z + 2q^T z \\
    s.t.& \ \ \ lb <= Az <= ub.
\end{split}
\end{equation}

The objective function we wish to minimize is stated in Equation \ref{eq:mpc_problem}. We present two equivalent formulations of the optimization problem. In order to differentiate the methods, we refer to them based on the size of the optimization matrices $P$ and $q$ which result. The two methods will be referred to in this work as the large matrix formulation and the small matrix formulation.

\subsection{Large Matrix Formulation}
Perhaps the most common way to formulate the optimization for MPC is to choose $z = [x_0^T, x_1^T, ..., x_{T}^T, u_0^T, u_1^T, ..., u_{T-1}^T, 1]^T$ which allows us to rewrite Equation \ref{eq:mpc_problem} as

\begin{equation}
J = (z - z_{goal})^T\begin{bmatrix} 
        Q_{big} & 0 & 0\\
        0 & R_{big} & 0\\
        0 & 0 & 0
        \end{bmatrix}(z - z_{goal})
\end{equation}

where $Q_{big} = I_{nT}\otimes Q$, $R_{big} = I_{mT}\otimes R$, and $\otimes$ represents the Kronecker product of two matrices. The zeros in the above matrix are included because of the value 1 included in the $z$ vector. This value is necessary to include the constant disturbance term $w$ as will be shown hereafter. Choosing $P$ as the block matrix in the above equation, we can further simplify this as

\begin{equation}
\begin{split}
    J &= (z - z_{goal})^T P (z - z_{goal}) \\
      &= z^T P z - 2z_{goal}^T P z + z_{goal}^T P z_{goal}.
\end{split}
\end{equation}

We now recognize that the last term in this expression is a constant offset and has no effect on the solution of the optimization. Eliminating the constant term we may write

\begin{equation}
\begin{split}
    J &= z^T P z - 2z_{goal}^T P z \\
      &= z^T P z + 2q^T z
\end{split}
\end{equation}

where $q^T = - z_{goal}^T P$ in order to fit our objective function into the form of Equation \ref{eq:convex_solver_form}.

In order to enforce the dynamics constraints which are necessary for MPC by solving an optimization of the form in Equation \ref{eq:convex_solver_form}, we must define a matrix $A$, as well as lower bounds $lb$ and upper bounds $ub$ on $z$. Note that when $lb=ub$, the constraint becomes an equality constraint. Given the discrete linear states space matrices $A_d$, $B_d$, $w_d$ and the initial state $x_0$, a common way to incorporate the constraints given in Equation \ref{eq:mpc_problem} into the form of Equation \ref{eq:convex_solver_form} is to choose

\begin{equation}
\label{eq:large_matrix_constraints}
A = \begin{bmatrix} 
    -I & 0 & \dots & 0 & 0 & \dots & 0 & 0\\
    A_d & -I & & 0 & B_d & & 0 & w_d\\
    & \ddots & \ddots & \vdots &  &  \ddots &
    \vdots & \vdots\\
    0 & \dots & A_d & -I & 0 & \dots & B_d & w_d
    \end{bmatrix}
\end{equation}

and 
\begin{equation}
lb = ub = \begin{bmatrix} 
    -x_0 \\
    0 \\
    \vdots \\
    1
    \end{bmatrix}
\end{equation}
Constraints on the inputs may be enforced by appending these constraints the bottom of the $A$, $lb$, and $ub$ matrices. Assuming the only other constraints are bounds on inputs, this optimization formulation exactly represents the minimization outlined in Equation \ref{eq:mpc_problem} with an optimization over $n(T+1) + Tm + 1$ variables with $n(T+1) + mT + 1$ constraints. Because the resulting $P$, $q$, and $A$ matrices are large, sparse matrices, we refer to this formulation of the optimization as the large matrix formulation of MPC.

%\question{Maybe consider defining "sparse" as a matrix with tons of zeros? You use that term quite a few times in the following text.}

\subsection{Small Matrix Formulation}
Another common way to formulate the optimization for MPC involves writing the vector of states over the time horizon as a function of only the inputs and initial condition. Because we have assumed linear dynamics, the vector of states over the time horizon can be written as 

\begin{align}
    \begin{bmatrix} 
        x_1 \\
        x_2 \\
        \vdots \\
        x_{T}
    \end{bmatrix} \nonumber
    &= 
    \begin{bmatrix} 
        A_d x_0 + B_d u_0 + w_d\\
        A_d x_1 + B_d u_1 + w_d\\
        \vdots \\
        A_d x_{T-1} + B_d u_{T-1} + w_d
    \end{bmatrix} \\ \nonumber
    &= 
    \begin{bmatrix} 
        A_d x_0 + B_d u_0 + w_d\\
        A_d^2 x_0 + A_d B_d u_0 + A_d w_d + B_d u_1 + w_d\\
        \vdots \\
        A_d ^{T} x_0 + \sum_{i=0}^{T-1}\big[A_d^i(B_d u_{T-1-i} + w_d)\big]
    \end{bmatrix} \\ \nonumber
    &=
    \begin{bmatrix} 
        B_d & 0 & 0 & \dots\\
        A_d B_d & B_d & 0 &\dots\\
        & \vdots \\
        A_d^{T-1} B_d & A_d^{T-2} B_d &\dots & B_d
    \end{bmatrix}
    \begin{bmatrix} 
    u_0 \\
    u_1 \\
    \vdots \\
    u_{T-1}
    \end{bmatrix} \\
    &\hspace{50pt} +\begin{bmatrix} 
                A_d x_0 + w_d\\
                A_d^2 x_0 + A_d w_d + w_d\\
                \vdots \\
                A_d^{T} x_0 + \sum_{i=0}^{T-1} A_d^i w_d
            \end{bmatrix} \\
    &= S z + v
    \label{eq:small_matrix_sz_v}
\end{align}

% or

% \begin{equation}
%     \begin{bmatrix} 
%     x_1 \\
%     x_2 \\
%     \vdots \\
%     x_{T+1}
%     \end{bmatrix} = S z + v
% \end{equation}

where $z=[u_0^T, u_1^T, ..., u_{T-1}^{T}]^T$.

By rewriting the states at each time step as a linear function of our inputs, we can now rewrite the cost function as

\begin{equation}
\begin{split}
    J =& (Sz+v - x_{goal})^T Q_{big} (Sz+v - x_{goal}) \\
    &+ (z-u_{goal})^T R_{big} (z-u_{goal}) \\
    =& z^T(S^T Q_{big} S + R_{big}) z \\
    &+ 2 z^T (S^T Q_{big} v - S^T Q_{big} x_{goal} - R_{big} u_{goal}) \\
    &+ 2 v^T Q_{big} v - 2v^T Q_{big} x_{goal} + x_{goal} Q_{big} x_{goal} \\
    &- u_{goal}^T R_{big} u_{goal}
\end{split}
\end{equation}

Realizing again that the last four terms which do not contain the optimization design variable $z$ are a constant offset and do not affect the optimization result, we can simplify further to

\begin{equation}
\begin{split}
    J =& z^T(S^T Q_{big} S + R_{big}) z \\
    &+ 2 z^T (S^T Q_{big} v - S^T Q_{big} x_{goal} - R_{big} u_{goal}) \\
    =& z^T P z + 2z^T q \\
\end{split}
\end{equation}

where $P = S^T Q_{big} S + R_{big}$ and $q = (S^T Q_{big} v - S^T Q_{big} x_{goal} - R_{big} u_{goal})$. Note that in this formulation, the dynamics constraints are implicitly enforced in the cost function and it is in the form of Equation \ref{eq:convex_solver_form} needed for use with a convex solver.

Assuming the only other constraints are bounds on inputs, this optimization formulation exactly represents the minimization outlined in Equation \ref{eq:mpc_problem} with an optimization over $mT$ variables with $mT$ constraints. Because the resulting $P$, $q$, and $A$ matrices are smaller than in the other formulation presented, we refer to this formulation of the optimization as the small matrix formulation of MPC.

\subsection{Parameterization}
Because the parameterization introduced in Section \ref{mpc_parameterization_method} is linear, it can also be incorporated into the convex optimization methods described above. 

Given $p$ knot points and a horizon of $T$ time steps, and assuming that one point is placed at the beginning and one at the end of the trajectory, we can define the distance between knot points in units of time steps ($\Delta T$) as 

\begin{equation}
    \Delta T = \frac{T-1}{p-1}.
\end{equation}

At any time step $k$, the input $u$ can be written as a linear combination of at most two knot points $U_{idx1}$ and $U_{idx2}$:

\begin{equation}
    u = (1-c) U_{idx1} + (c) U_{idx2}.
\end{equation}
where 
\begin{equation}
\begin{split}
    idx1 &= floor(\frac{k}{\Delta T}) \\
    idx2 &= idx1 + 1 \\
    c &= \frac{k - (idx1) \Delta T}{\Delta T}.
\end{split}
\end{equation}

The dynamics can then be written as

\begin{equation}
    x_{k+1} = A_d x_k + (1-c) B_d U_{idx1} + (c)B_d U_{idx2} + w_d.
\end{equation}

In order to parameterize both the large matrix formulation or the small matrix formulation of MPC, we include the knot points ($U$) in the vector of design variables ($z$) instead of the inputs at each time step ($u$). Other matrices in the optimization must also be modified. Although the modifications to these matrices are not complex, a general form for any horizon and number of parameters quickly becomes difficult to understand. Instead, we provide simple examples for the specific case of a horizon of $T=5$ time steps using $p=3$ knot points.

\subsection{Parameterized Large Matrix Formulation}

For the large matrix formulation of MPC, in addition to the vector of design variables ($z$), the constraint matrix $A$ must be modified. Instead of a matrix of the form described in Equation \ref{eq:large_matrix_constraints}, for the case of $T=5$ and $p=3$ we obtain the matrix

\begin{equation}
\begin{split}
&A = \\
    &\begin{bmatrix} 
    -I & 0 & 0 & 0 & 0 & 0 & 0 & 0 & 0 & 0\\
    A_d & -I & 0 & 0 & 0 & 0 & B_d & 0 & 0 & w_d\\
    0 & A_d & -I & 0 & 0 & 0 & \frac{1}{2}B_d & \frac{1}{2}B_d & 0 & w_d\\
    0 & 0 & A_d & -I & 0 & 0 & 0 & B_d & 0 & w_d\\
    0 & 0 & 0 & A_d & -I & 0 & 0 & \frac{1}{2}B_d & \frac{1}{2}B_d & w_d\\
    0 & 0 & 0 & 0 & A_d & -I & 0 & 0 & B_d & w_d
    \end{bmatrix}.
\end{split}
\end{equation}

The only change to this matrix is in the portion containing the matrix $B_d$ and involves the linear combination of at most two knot points for each input.

Assuming the only constraints are bounds on inputs, the parameterized large matrix formulation of MPC requires optimization over $n(T+1) + mp + 1$ variables with $n(T+1) + mp + 1$ constraints. This represents a reduction in the number of optimization design variables and constraints by $m(T+1-p)$.

\subsection{Parameterized Small Matrix Formulation}
For the small matrix formulation, in addition to the vector of design variables ($z$), the matrix $S$ in Equation \ref{eq:small_matrix_sz_v} must be modified which in turn affects the matrix $P$. Note that the matrix $v$ in Equation \ref{eq:small_matrix_sz_v} does not change at all. For the parameterized $S$ matrix in the case of $T=5$ and $p=3$ we obtain

\begin{equation}
\begin{split}
&S = \\
    % &\begin{bmatrix} 
    %     B_d & 0 & 0\\
    %     A_d B_d + B_d(\frac{1}{2}) & B_d(\frac{1}{2}) & 0\\
    %     A_d^2 B_d + A_d B_d (\frac{1}{2}) & A_d B_d(\frac{1}{2}) + B_d(\frac{1}{2}) & B_d(\frac{1}{2})\\
    %     A_d^3 B_d + A_d^2 B_d (\frac{1}{2}) & A_d^2 B_d(\frac{1}{2}) + A_d B_d(\frac{1}{2}) & A_d B_d(\frac{1}{2}) + B_d(\frac{1}{2})\\
    % \end{bmatrix} \\
    &\begin{bmatrix} 
        B_d & 0 & 0\\
        (A_d + \frac{1}{2}I)B_d & \frac{1}{2}B_d & 0\\
        (A_d^2 + \frac{1}{2}A_d)B_d & (\frac{1}{2}A_d + I)B_d & 0\\
        (A_d^3 + \frac{1}{2}A_d^2) B_d  & (\frac{1}{2}A_d^2 + A_d + \frac{1}{2}I) B_d & \frac{1}{2}B_d \\
        (A_d^4 + \frac{1}{2}A_d^3) B_d  & (\frac{1}{2}A_d^3 + A_d^2 + \frac{1}{2}A_d) B_d & (\frac{1}{2}A_d + I) B_d \\
    \end{bmatrix}    
\end{split}
\end{equation}
where $I$ denotes the identity matrix.

A simple way to calculate this matrix is to start from the top and work downwards. Each progressive row is simply the previous row pre-multiplied by $A_d$, then summed with $B_d$ in the correct proportions in each column.

The parameterized small matrix formulation of MPC requires optimization over $mp$ variables with $mp$ constraints. The difference in the number of design variables and constraints in this case is also $m(T+1 - p)$. Although this reduction in the number of variables and constraints is identical to the reduction for the large matrix formulation, because the small matrix formulation started with fewer variables and constraints it represents a proportionally larger reduction in the optimization size.

\subsection{Solve Times of Parameterized Convex Solver MPC - Experimental Setup}
\label{convex_solver_experimental_setup}

Because we are interested in how our methods scale to high DoF robots, we choose to perform experiments with simulated robots with between one and thirteen links. We refer to these robots as N link robots. Each link is assumed to be the same mass (1 kg) and length (.25m) and each robot's workspace is assumed to be in the plane perpendicular to the gravity vector. The dynamic equations take the canonical form 
\begin{equation}\label{eq:threelink}
    M(q)\Ddot{q} + C(q,\dot{q}) + b\dot{q} = \tau
\end{equation}
where $q$ is the vector of generalized coordinates, $M(q)$ is a configuration dependent inertia matrix, $C(q,\dot{q})$ represents torques produced by centrifugal and Coriolis forces, $b$ is a viscous damping coefficient (.01), and $\tau$ are applied torques from the motors.

The main advantages of MPC are that it easily incorporates constraints and performs an optimization over a future time horizon, leading to less ``greedy" behavior. In order to test our methods in the most useful scenarios we therefore choose a damping coefficient $b$ such that the robots are all underdamped and enforce a torque constraint of two Nm at each joint. This poses a more interesting and challenging problem than that of an over-damped robot or one with infinite available torque.

Experiments are performed to quantify the effects of different MPC formulations, as well as parameterization, on MPC solve times.
%Because the different formulations are solving the exact same problem, MPC performance is identical.
We use a state of the art convex solver (OSQP) \cite{osqp} to perform the optimization. We record both the run time reported by the solver and the total MPC solve time, which includes the time needed to calculate the matrices $P$, $q$, $A$, $lb$, and $ub$. We report these solve times as the optimization solve time and MPC solve time respectively. Because the performance of parameterized methods was already explored in Section \ref{parameterized_mpc_performance_setup}, we do not attempt to quantify performance in these experiments.

The first experiment is performed by choosing random initial and goal states for a robot, calculating the dynamics matrices for the robot based on initial conditions, and then using each MPC solver to solve for the next input to apply. Because we are only interested in solve times, the input is never actually applied to any system, but the process of sampling random initial and goal conditions is repeated 100 times. This process is carried out for robots with one to thirteen links. A horizon of 50 time steps is used for every solve, and the parameterized methods used five parameters. Median optimization solve times and MPC solve times for each MPC formulation and each number of links can be seen in Figures \ref{fig:horizon_50_median_solve_times} and \ref{fig:horizon_50_median_total_times} respectively.

The second experiment carried out is exactly the same as the first, but with a horizon of 100 instead of 50. The parameterized methods still use five parameters. Median optimization solve times and MPC solve times for each MPC formulation and horizon are reported in Figures \ref{fig:horizon_100_median_solve_times} and \ref{fig:horizon_100_median_total_times} respectively.

\subsection{Solve Times of Parameterized Convex Solver MPC - Experimental Results}

\begin{figure}
    \centering
    \includegraphics[width=\linewidth]{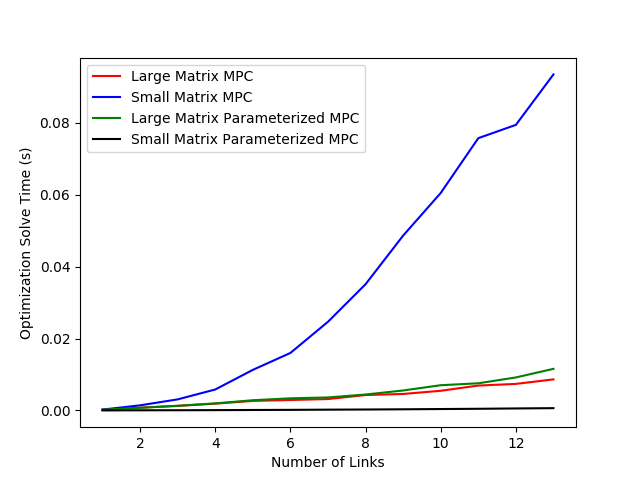}
    \caption{Median optimization solve times over hundreds of trials for a horizon of 50 time steps.}
    \label{fig:horizon_50_median_solve_times}
\end{figure}

\begin{figure}
    \centering
    \includegraphics[width=\linewidth]{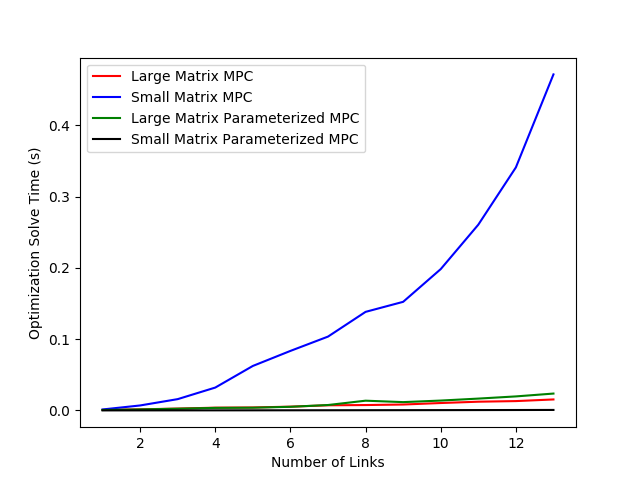}
    \caption{Median optimization solve times over hundreds of trials for a horizon of 100 time steps.}
    \label{fig:horizon_100_median_solve_times}
\end{figure}

\begin{figure}
    \centering
    \includegraphics[width=\linewidth]{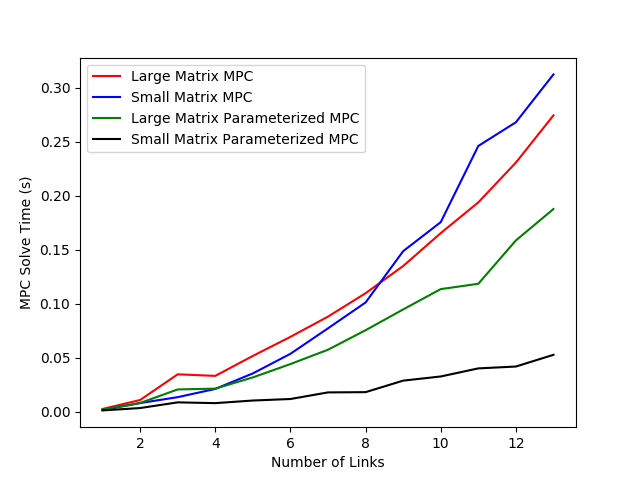}
    \caption{Median MPC solve times over hundreds of trials for a horizon of 50 time steps.}
    \label{fig:horizon_50_median_total_times}
\end{figure}

\begin{figure}
    \centering
    \includegraphics[width=\linewidth]{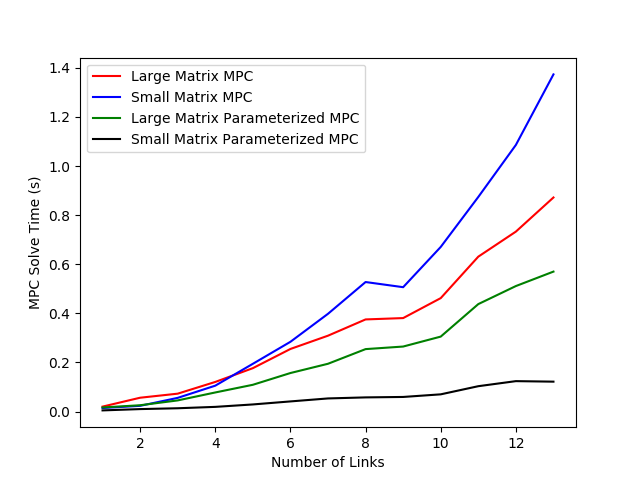}
    \caption{Median MPC solve times over hundreds of trials for a horizon of 100 time steps.}
    \label{fig:horizon_100_median_total_times}
\end{figure}

Looking first at just the optimization solve times in Figures \ref{fig:horizon_50_median_solve_times} and \ref{fig:horizon_100_median_solve_times} we note the expected result that the solve times for robots with more links are higher than those with fewer links. As can be seen in both figures, the small matrix formulation of MPC solves very quickly for small problem sizes, but seems to scale poorly to larger problem sizes. Perhaps surprisingly, even though the number of variables and constraints is larger for the large matrix formulation, it solves faster than the small matrix formulation for large problems.

This counter-intuitive result may be explained by the optimization method used. Many fast convex solvers depend on a factorization of the $P$, $A$, and $q$ matrices from Equation \ref{eq:convex_solver_form} to solve a system of linear equations. Very efficient solution methods exist for performing these linear system solves for large, sparse matrices. By decreasing the search space in the small matrix formulation we have exchanged large sparse matrices for small dense matrices.

A similar phenomenon is observed with the parameterization of the large matrix formulation of MPC. We see that for small problem sizes, parameterization of the large matrix formulation has little effect on the solve time. In fact if anything, shrinking the search space of the large matrix formulation seems to have the effect of increasing the solve time. Again, we believe this is related to the linear system solution of a sparse matrix that we make more dense through parameterization.

The parameterized small matrix formulation of MPC consistently has the lowest optimization solve times of any MPC formulation. Although the matrices used in the parameterized small matrix formulation are very dense, they are also very small. Although it is difficult to tell because of the scale of the plots, the parameterized small matrix MPC optimization solve times in Figures \ref{fig:horizon_50_median_solve_times} and \ref{fig:horizon_100_median_solve_times} are near identical, despite the length of the horizon doubling.
When we remember that for the small matrix formulation, the optimization problem size is not a function of horizon length, but only number of parameters, this result makes sense. If the number of parameters is held constant, the horizon can be made arbitrarily large with no effect whatsoever on solve time. The same is not true however, for the time taken to calculate the matrices $P$ and $q$.

By looking at Figures \ref{fig:horizon_50_median_total_times} and \ref{fig:horizon_100_median_total_times} we can see the total MPC solve times. These solve times include the optimization solve time plus the time taken to calculate the matrices $P$, $Q$, $A$, $lb$, and $ub$. We recognize that these results may vary based on specific methods and code implementations for calculating these matrices, however we submit that the trends presented represent trends dictated by the size and complexity of the matrices. For the large matrix formulations, most of this extra time is taken to form the large constraint matrix $A$ which contains the dynamics constraints. 
%Because the dynamics constraint is implicitly enforced in the objective function 
For the small matrix formulations, there is greater time spent calculating the matrices $P$ and $q$.

For both the horizon of 50 and the horizon of 100, we can see that the overall effect of parameterization was a reduction in the MPC solve time. This is true even for the large matrix formulation, where the optimization solve time is not decreased by parameterization. This is likely because the matrices used in the parameterized version are much smaller and take less time to form before optimization.

Apart from the trend that parameterization decreases MPC solve time for either formulation of MPC, we can see that parameterization has a greater effect on the MPC solve time of the small matrix formulation, especially for long horizons and complex systems. We see that for robots with many links, the small matrix formulation has the longest MPC solve times, while the parameterized version has the shortest MPC solve times. We also note that the parameterized small matrix formulation seems the least sensitive to the number of links. This is evidence that this method scales better to high DoF systems than the others.%traditional MPC.

%Although the optimization solve times vary dramatically, it is interesting to note that the total MPC solve times are generally closer. This represents the fact that 

%For the small matrix formulation we expect and in fact find that by increasing the number of parameters increases the solve time up to limit of the non-parameterized version. The very low solve times for the parameterized small matrix formulation can only be explained by the fact that the optimization has been reduced to a very small problem. Just as the small matrix formulation solved faster than the large matrix formulation for the one link robot with short horizons (Figure \ref{fig:onelink_parameterized_median_solve_times}), if the optimization problem is kept small enough then the fact that the matrices are dense seems not to matter as much.

These results combined with those found in Section \ref{parameterized_mpc_performance_results}, strongly suggest that when performing MPC for high DoF systems, instead of decreasing horizon length to satisfy demands on solve time, we should instead reformulate and parameterize the input trajectory to decrease solve times. We have established that this method leads to faster solve times with little to no loss in performance. In fact, these results suggest that most traditional MPC controllers can be run at a higher rate through parameterization.

While the results found for this section are found using one specific solver, we expect that similar results would be found with other fast solvers which utilize factorization of sparse matrices for finding the solutions to a system of linear equations.

\section{Evolutionary MPC}
\label{empc}

Up to this point we have discussed the parameterization of MPC, its effects on MPC performance and robustness, and how to implement parameterized MPC for use with a convex solver. We have established that MPC solve times can be decreased drastically by using the small matrix formulation with parameterization.

We now turn our attention to another method of implementing parameterized MPC which is parallelizable, scalable, and can admit much more flexible cost and dynamics constraints. Because this method utilizes an evolutionary algorithm to perform optimization, we refer to this method as evolutionary MPC (EMPC)\footnote{We are grateful to Dr. Greg Stewart from The University of British Columbia for suggesting the random sampling of the configuration/state space.} and we implement it on a GPU. This is similar to the work in \cite{hyatt2017real} where preliminary work with EMPC applied it to a seven DoF compliant robot arm. The current work builds on \cite{hyatt2017real} by relaxing heuristic assumptions about the knot points, using a linear parameterization, and simplifying the evolutionary algorithm.

\begin{figure}[hbt]
  \centering
  \includegraphics[width=1.0\linewidth]{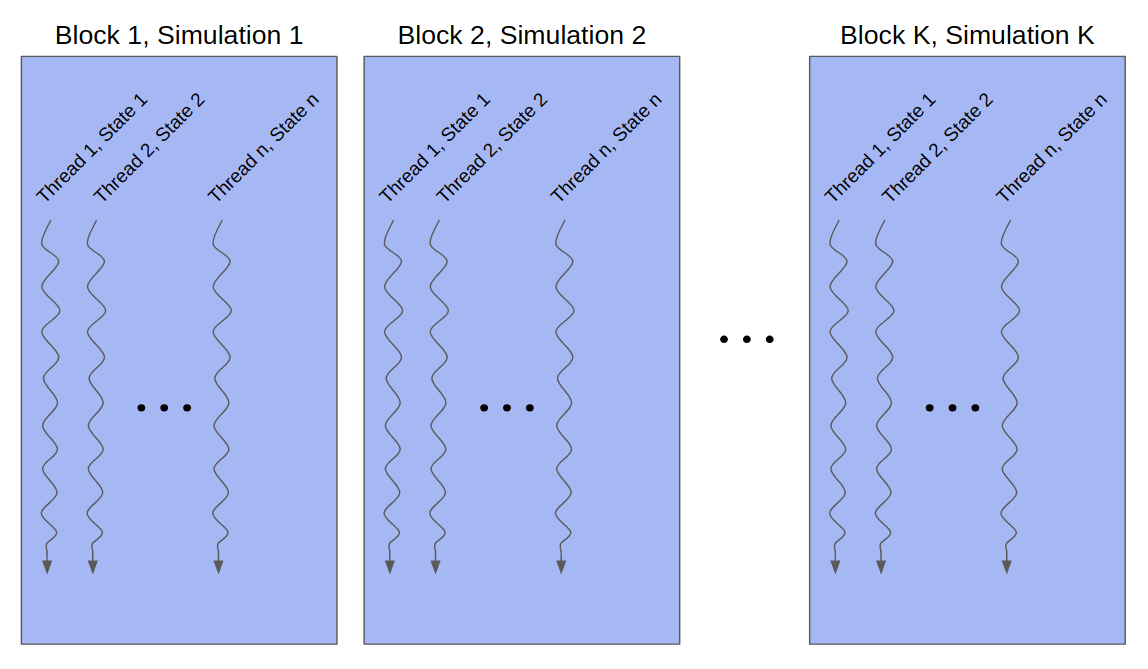}
  \caption{Graphical explanation of CUDA programming hierarchy as implemented in this paper. One simulation is launched in each block and numSims blocks are launched simultaneously. After all blocks have finished, each block/simulation has calculated a cost associated with an input trajectory.}
  \label{fig:cuda_heirarchy}
\end{figure}

%\subsection{Evolutionary MPC on a GPU}
GPUs are designed to launch thousands of threads which all perform the same function on different data. In the CUDA programming interface these threads are assigned unique indices and are organized in structures called blocks as seen in Figure \ref{fig:cuda_heirarchy}. This hierarchy is important to understand because only threads within the same block can be made to run synchronously and may share common memory. These considerations are important when efficiently simulating and evaluating costs in parallel on a GPU.

At a high level, EMPC utilizes the parallel computing capability of GPUs by evaluating the fitness of hundreds or thousands of input trajectories simultaneously. The best of these trajectories are then mated and mutated to produce a new generation of trajectories which should theoretically be better than the last. As in typical MPC, only the first input of the best trajectory is applied to the actual system. Unlike typical MPC however, EMPC does not wait for any convergence criteria before applying the next input to the system. Instead, the best input is applied to the system after each generation. Despite applying what are almost certainly sub-optimal inputs, because EMPC is continuously improving its solution as it progresses towards the goal, EMPC performs similarly to MPC using a state of the art convex solver, while allowing for shorter solve times and longer horizons (see Section \ref{simulation_experiments}).

%The success of EMPC in finding good solutions is owed in part to the parameterization of the input space. Instead of optimizing over every input at every time step (a very large dimensional search space for high DoF long horizon problems), we instead represent a time varying input trajectory as a function of time and a small number of parameters. How to best choose this parameterization is the topic of Section \ref{mpc_parameterization}, but a few intuitive examples are splines or linear combinations of basis polynomials. Choosing such representations makes the dimensionality of the search space invariant with respect to horizon length, since the trajectory can always be defined by a set number of knot points or polynomial coefficients.

Because evolutionary algorithms are gradient-free and non-local optimizations, the type of parameterization used is very flexible. For the sake of generality, for this section we will let $\mathbf{U}$ represent the parameters which define all system inputs over a trajectory, and $\mathbf{U}(i)$ represent the parameters which represent the $i$th input over the trajectory. We will let $get\_u\_from\_U(\mathbf{U}(i),t)$ represent the function which maps from a set of parameters and a time to the actual input applied to the system at that time.

The EMPC algorithm is highly parallelizable, especially because of the assumption of linear dynamics. In order to more fully exploit the inherent parallelism of the problem using the CUDA programming interface for GPUs, we organize our calculations as seen in Figure \ref{fig:cuda_heirarchy}. Each simulation is carried out within one block and each thread within that block is assigned one system state. The threads with $threadIdx < m$ are assigned one input each as well. This assumes that the number of states is equal to or greater than the number of inputs.

The computations for one thread are outlined in Algorithm \ref{alg:EMPC} lines 3-17 for reference, and are discussed more in detail below. At the beginning of each generation, in the case of a cold start, the parameters $\mathbf{U}$ are selected from a random uniform distribution. In the case of a warm start, these parameters are produced by mating two sets of randomly chosen ``parent" parameters. This mating is accomplished by looping through each parameter in the trajectory and giving the child 50\% probability of inheriting each parameter from each parent. This process of crossover allows the best trajectories to be combined with the hope of finding new trajectories which are the best parts of both parents.
%, however if all of the parents are similar and no other mutation is allowed it may result in stagnation.

In order to encourage exploration and avoid the tendency of populations to stagnate in local optima, we introduce changes in the input population with mutations. After mating is finished, each input has a 50\% probability of being mutated. If mutated, additive noise sampled from a mean-zero normal distribution with standard deviation $\sigma_{noise}$ is added to the parameters.

The number of parents ($numParents$),  the number of simulations $numSims$, and the exploration noise $\sigma_{noise}$ are tune-able parameters of the algorithm. If $threadIdx >= m$, then the thread waits for the first $m$ threads to reach line 11 of Algorithm 1 before continuing.

\begin{algorithm}
   \caption{Evolutionary Model Predictive Control}\label{alg:EMPC}
    \begin{algorithmic}[1]
    \For{each Simulation (Parallel block)}
        \For{$i = 0$ to $n$ (Parallel thread)}
            \If{Cold Start}
                \State $\mathbf{U}(i) \sim \mathcal{U}(\mathbf{u_{min}}(i),\mathbf{u_{max}}(i))$
            \ElsIf{Warm Start}
                \State $P1 \sim \mathcal{U}(0,numParents)$
                \State $P2 \sim \mathcal{U}(0,numParents)$
                \State $Noise \sim \mathcal{N}(0,\sigma_{noise})$
                \State $\mathbf{U}(i) = mate(\mathbf{U_{P1}}(i),\mathbf{U_{P2}}(i)) + Noise$
            \EndIf
            \For{$t = 0$ to ${T}$}
                \State $\mathbf{u}(i) = get\_u\_from\_U(\mathbf{U}(i),t)$
                \State Synchronize $\mathbf{x}$ and $\mathbf{u}$ across block
                \State $\mathbf{x}(i) = A(i,:)\cdot \mathbf{x} + B(i,:) \cdot \mathbf{u} + \mathbf{w}$
                \State $J_i += stage\_cost(\mathbf{x}(i),\mathbf{u}(i),\mathbf{x_{goal}},\mathbf{u_{goal}})$
            \EndFor
            \State $J_i += terminal\_cost(\mathbf{x}(i),\mathbf{u}(i),\mathbf{x_{goal}},\mathbf{u_{goal}})$
        \EndFor
        \State $J = \sum_{i=0}^{n} J_i$
    \EndFor
    \State $\mathbf{U_0, U_1, ..., U_{numParents}} = \mathbf{U}$'s with lowest J's
    \For{$i = 0$ to $m$}
        \State $\mathbf{u^*}(i) = get\_u\_from\_U(\mathbf{U_{best\_parent}}(i),0)$
    \EndFor
\end{algorithmic}
\end{algorithm}

%\begin{figure}[hbt]
%  \centering
%  \includegraphics[width=1.0\linewidth]{figures/gpu_ga_mpc_flowchart.png}
%  \caption{Flowchart for the execution of EMPC on a GPU}
%  \label{fig:empc_flowchart}
%\end{figure} 

\subsection{Simulation and Cost Calculation}
The simulation of an input trajectory over a horizon is done sequentially using Equation \ref{eq:first_order_integration}. Because of the coupling between states, the calculation of each element of $\mathbf{x_{i+1}}$ depends on the entire vector of states $\mathbf{x_i}$ as well as the entire vector of inputs $\mathbf{u_i}$. Because threads in the same block can access the same shared memory and can be made to synchronize at certain points, each thread can calculate its own state in the vector $\mathbf{x_{i+1}}$ using the states and inputs saved in shared memory. Then after all states have performed this calculation, the states and inputs in shared memory can be updated again.

Some of EMPC's greatest strengths come from this approach to simulation. Although the state space dynamics are linear, because we calculate each state and input at each timestep, we can include some nonlinear effects such as
%saturation on inputs, state constraints, 
hysteresis, contact, and impacts. These types of dynamic effects can simply be placed in if/else statements during simulation. Another benefit of this type of simulation is that while convex solvers may find a problem infeasible and return no solution at all, EMPC can be made to always return the best possible solution, and the constraints describing the dynamics of the system are satisfied implicitly as part of the simulation.

At each time step in this forward simulation, each thread can evaluate the stage cost associated with its state as well as a terminal cost at the final time step. The total cost is updated at each time step, then is summed with the rest of the costs from the other threads and is saved to device memory as the simulation cost. Because we are performing a gradient-free optimization, this allows almost unlimited flexibility in the type of cost function used. Cost functions need not be convex or even continuous, so they may contain if/else statements, piecewise functions, and other functions.

%\todo{Now that we're just doing 1 sim per block maybe all of the state costs should be stored in shared memory. Then the thread with threadIdx0 can just sum them up real quick and put it in dout...I should test this and see if it is faster than what I currently do. It would also save some GPU memory copying time and space.}

Once all simulations are finished, we are left with $numSims$ $\mathbf{U}$ arrays, each with an associated cost. Parent selection simply consists of taking the $numParents$ $\mathbf{U}$ arrays with lowest associated costs and preserving them for the next generation. In order to find $u^*$ (the optimal $u$) for use with MPC, we take the $\mathbf{U}$ array with lowest associated cost, and find the $u(i)$ defined by $get\_u\_from\_U(\mathbf{U_{best\_parent}}(i),t=0)$ for each input $i$. This corresponds to the inputs applied at the first time step given the $\mathbf{U}$ array from the lowest cost simulation.

\section{Comparing Evolutionary, Parameterized Convex Solver, and Traditional MPC}
\label{comparing_mpc_methods}

\subsection{Simulation Experiments - Setup}
\label{simulation_experiments}

%\todo{Add small matrix parameterized MPC to the experiment and take out different CPUs. Also compare optimization and MPC solve times separately like in the previous section.}

In this section experiments are performed in order to compare the solve times and performance of EMPC to traditional and parameterized gradient-based MPC. We are especially interested in how these different solution methods scale to high DoF problems. We again use OSQP as the gradient-based solver in MPC. We perform experiments for this section using the same simulated N link robots described in Section \ref{convex_solver_experimental_setup}.

To test the performance of each MPC controller, we simulate using the controllers to perform control at 100 Hz for ten seconds. Performance is measured using the same ``actual cost" metric as in Section \ref{parameterized_mpc_performance_setup}, where ``actual cost" is calculated by evaluating the cost function used in MPC over the ten second simulation. The same exact cost function and weightings are used for both MPC methods and control is simulated to be done at 100 Hz, even when solve times exceed .01s. The parameterized versions of MPC are using only three knot points. The horizon used for MPC is 100 time steps with each step representing .01s. Each MPC method is used to control from an initial position at rest, to a goal position at rest. To avoid any bias from a particular part of the robot's workspace, we again ran several trials with initial and goal positions sampled from a uniform random distribution and report statistics over all trials.

Solve times are recorded for each MPC solve in every trial. The optimization solve time recorded for MPC utilizing OSQP corresponds to the ``run time" reported by the OSQP solver. This includes time taken to update the sparse matrices within the solver and solve the optimization. Likewise, the optimization solve time recorded for EMPC includes the time taken to copy needed data to the GPU, perform one or more iterations of the genetic algorithm, and copy needed information back from the GPU. Time taken to calculate the dynamics matrices was not included in either solve time. The OSQP MPC solver also required that the dynamics matrices be formatted into large constraint and cost matrices which are then fed into the solver, while for EMPC this step is not necessary. The time taken to construct these matrices plus the optimization solve time is reported as the MPC solve time.

In order to examine the scalability of both traditional MPC and EMPC, we perform experiments on robots with one to thirteen links. We report both performance and solve time statistics as a means of measuring scalability. Because EMPC is parallelized for use on a GPU, we also ran experiments on two different GPUs (an NVIDIA GeForce GTX 750 Ti and an NVIDIA GeForce GTX Titan X). This comparison is interesting because the Titan X contains many more parallelized processing cores than the 750 Ti (3072 vs 640). These experiments help demonstrate what effects GPU improvement may have on parallelized MPC methods.

%It is important to note that in practice, if solve times exceed time dictated by the control rate (in this case .01s), either the controller would have to be run at a slower rate, or the horizon length would have to be decreased until fast enough solve times can be achieved. Both options generally lead to worse performance. 
%\todo{Note that this happens more to OSQP, if not only to OSQP}
%While slower solve times had no effect on control rate or performance for these simulation experiments, the experiments performed in hardware in Section \ref{hardware_experiments_setup} of necessity demonstrate the effect of solve times on control rate and performance.
We also perform experiments using a variant of EMPC which performs three iterations of the evolutionary algorithm before returning a value for use with MPC. This should allow the evolutionary algorithm to find a better solution at the cost of longer solve times.

An example of the joint angle response of a six link robot for these trials is demonstrated in Figure \ref{fig:sample_trajectory}.

\subsection{Simulation Experiments - Results}

% \begin{figure}[hbt]
%   \centering
%   \includegraphics[width=1.0\linewidth]{figures/sample_trajectory.png}
%   \caption{}
%   \label{fig:sample_trajectory}
% \end{figure} 

\begin{figure*}[hbt]
  \centering
  \includegraphics[width=1.0\textwidth]{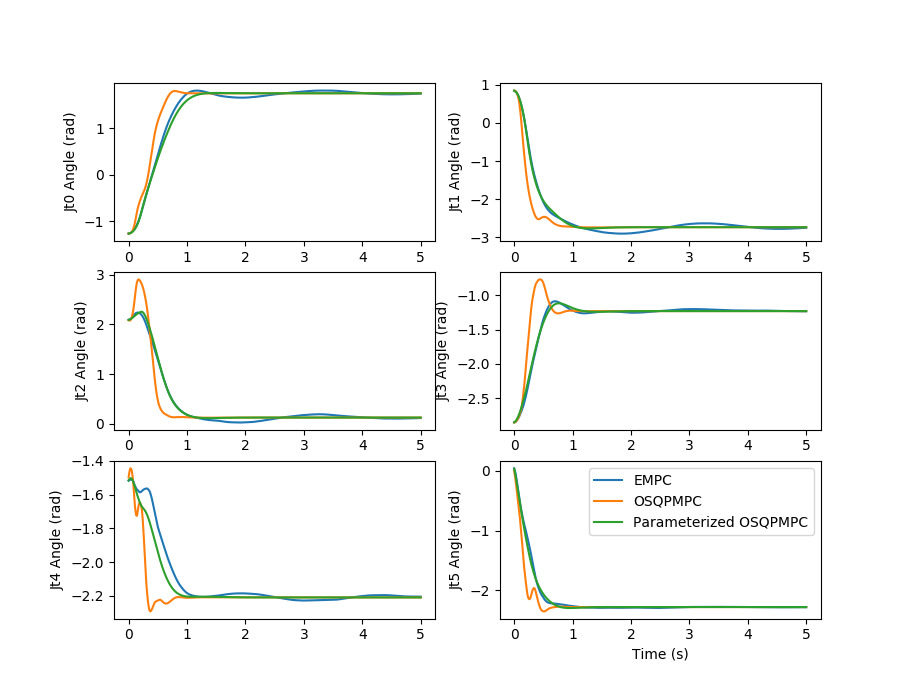}
  \caption{Sample joint angle response for a six link planar robot controlled by three different Model Predictive Controllers. Note that the parameterized methods (blue and green) are more conservative and produce less overshoot.}
  \label{fig:sample_trajectory}
\end{figure*} 

\begin{figure}[hbt]
  \centering
  \includegraphics[width=1.0\linewidth]{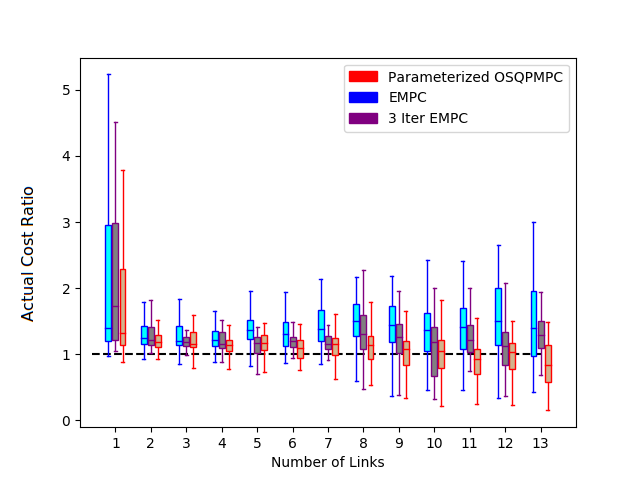}
  \caption{Ratio of actual cost using EMPC and a parameterized gradient-based solver to traditional MPC. The dashed line at one denotes performance equal to traditional MPC, below one is better, and above is worse.}
  \label{fig:lempc_vs_osqp_performance}
\end{figure}

A direct comparison of performance between parameterized convex MPC and EMPC, using traditional MPC as a baseline, can be seen in Figure \ref{fig:lempc_vs_osqp_performance}. We report the ratio of actual cost using each controller to the actual cost using traditional MPC. A ratio of one (the dashed line) corresponds to the same level of performance as traditional MPC. Consistent with the findings in Section \ref{parameterized_mpc_performance_results}, the performance of MPC with parameterization is generally not as good as that of traditional MPC for simple systems. The increased cost for any robot except the one link however is generally pretty small, accruing less than 20\% more cost when using the convex solver. For higher DoF systems however, the convex parameterized MPC accrues lower costs than traditional MPC. This is likely because of the conservative nature of parameterized MPC as outlined in Section \ref{parameterized_mpc_robustness_results}. A clear example of the conservative nature of the parameterized MPC approach is shown in Figure \ref{fig:sample_trajectory} for a six link robot. It can be seen that the un-parameterized MPC method has a faster rise time, but more overshoot.

The performance of EMPC is slightly worse than that of the convex solver, but again, the difference is not very large. It is interesting to note that by performing more iterations or ``generations" for the evolutionary algorithm, the EMPC cost begins to approach that of the parameterized convex solver. This essentially represents the classical trade-off between speed and quality of solution.

%The performance of EMPC is not surprisingly even slightly worse. Despite using the same parameterization as the parameterized gradient-based solver, EMPC only carries out one generation of an evolutionary algorithm for each MPC solve instead of an entire gradient-based optimization. However, there is an interesting downward trend in the cost ratio with increasing number of links which suggests that for more complex systems the performance of both EMPC and parameterized gradient-based MPC is more comparable to that of traditional MPC. The actual cost ratio of parameterized gradient-based MPC even dips below one for more complex systems, meaning that performance was better with parameterization than without for large DoF nonlinear systems.

\begin{figure*}[hbt]
  \centering
  \includegraphics[width=1.0\textwidth]{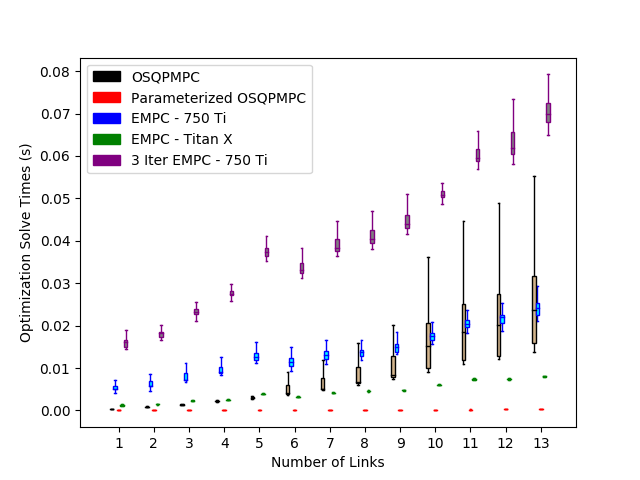}
  \caption{Comparison of optimization solve times for increasingly complex robots}
  \label{fig:lempc_vs_osqp_solve_times}
\end{figure*}

Figure \ref{fig:lempc_vs_osqp_solve_times} represents the relationship between the number of robot links and the optimization solve times for all of the MPC controllers. We see the same trend in solve times for the gradient-based solvers as in Figures \ref{fig:horizon_100_median_solve_times} and \ref{fig:horizon_50_median_solve_times}, however the median solve times in these experiments are much lower. That is because in these experiments the solvers are allowed to warm start themselves with a previous solution, which was often near the optimum once the robot reached a steady state equilibrium. The fact that each of our simulations lasts ten seconds means that a great deal of the solve times recorded are from this steady state regime.

For low DoF systems, EMPC optimization solve times are greater than both gradient-based solvers. However we see in Figure \ref{fig:lempc_vs_osqp_solve_times} that one iteration EMPC using either of the GPUs we presented has faster optimization solve times than traditional MPC for ten or more links. By increasing the number of iterations in EMPC by a factor of three we seem to have also increased the solve time by about the same factor. 

It is also interesting to note the improvement in solve time by changing only the GPU. It seems that by using the larger GPU, the solve time is decreased by about three ms, but this three ms decrease does not seem affected by the number of links. It is not unreasonable to assume that by using a more powerful GPU (such as the Titan V) solve times would be further reduced. Reductions in speed could also be achieved by using multiple GPUs. This scalability to high DoF problems through parallel computation is one of the main benefits of the EMPC approach.

\begin{figure*}[hbt]
  \centering
  \includegraphics[width=1.0\textwidth]{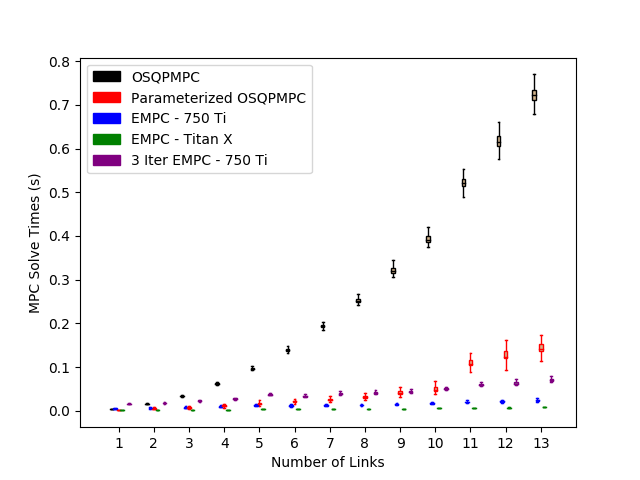}
  \caption{Comparison of MPC solve times for increasingly complex robots}
  \label{fig:lempc_vs_osqp_mpc_solve_times}
\end{figure*}

Figure \ref{fig:lempc_vs_osqp_mpc_solve_times} shows the MPC solve times for each MPC controller for a varying number of robot links. Again, consistent with the results in Section \ref{parameterized_mpc_performance_results}, we find that traditional MPC scales poorly with increasing robot complexity. We also confirm the result that the parameterized gradient-based MPC increases the tractability of MPC. The most impressive feature of these results however is how EMPC scales with increasing DoF of the robot. Because EMPC does not require the formation of large matrices for definition of the cost function or constraints, the MPC solve time is identical to that of the optimization solve time. This demonstrates the fact that the EMPC method scales very gracefully to higher DoF systems.

We see from these results an expected trade-off between tractability and the quality of the solution. Both methods presented in this work are more tractable than traditional MPC and provide high quality solutions. For high DoF systems, one iteration EMPC has the fastest solution time by far, however it also has incurs the highest cost. By increasing the number of ``generations" within an EMPC solve the cost can be decreased, but with slightly higher solve times. Parameterized convex solver MPC provides the lowest cost solutions (even lower than traditional MPC for the high DoF cases), but has a higher solve time for high DoF cases.

Because of the tractability and performance trade-off which exists between parameterized convex solver MPC and EMPC, it is difficult to say that one will be more tractable or provide better performance than the other in all cases. However Figure \ref{fig:lempc_vs_osqp_mpc_solve_times} of MPC solve times and Figure \ref{fig:lempc_vs_osqp_performance} of performance provide convincing evidence that control trajectory parameterization can make MPC far more tractable with very little effect on performance. This is an important result which may enable MPC for many robotic systems for which it was not previously possible.

% \begin{figure}[hbt]
%   \centering
%   \includegraphics[width=1.0\linewidth]{figures/mpc_solve_times_n_link_zoomed.png}
%   \caption{}
%   \label{fig:lempc_vs_osqp_solve_times_zoomed}
% \end{figure} 

\subsection{Hardware Experiments - Setup}
\label{hardware_experiments_setup}

\begin{figure}[hbt]
  \centering
  \includegraphics[width=\linewidth]{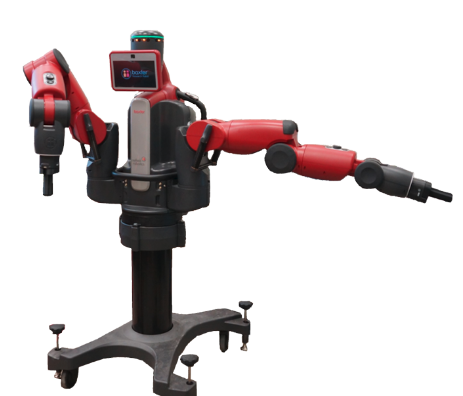}
  \caption{Seven DoF Baxter robot used for the hardware experiments}
  \label{fig:baxter}
\end{figure}

In order to demonstrate that both EMPC and parameterized gradient-based MPC are viable approaches to performing high DoF MPC for real systems, we implement them both on hardware. Specifically, we implement both for position control of one seven DoF arm of the Baxter robot seen in Figure \ref{fig:baxter}. A PD controller on position was run at 500 Hz and tuned with very low gains ($K_p$, $K_d$) in order to produce an under-damped system which is more safe to operate near humans or delicate equipment, but harder to control for smooth motion. The inputs selected by MPC are set points in joint space ($q_{des}$) for the PD controller which then applies torques directly to the robot.

We model the kinematics and dynamics of the robot using parameters provided by the manufacturer for lengths and masses. The robot automatically applies torques to oppose those of gravity, so we choose to model the robot as if there are no gravity torques. The full dynamics of the arm are 

\begin{equation}\label{eq:blow_molded}
    M(q)\Ddot{q} + C(q,\dot{q}) = K_{p}(q_{des}-q) - K_{d}\dot{q}
\end{equation}

where $q$ is the vector of joint angles, $M(q)$ and $C(q,\dot{q})$ are the inertia matrix and Coriolis and centrifugal terms respectively. $K_p$ and $K_d$ are the PD control gains and $q_{des}$ is a vector of inputs to the system. 

In the hardware experiment we send several step commands which involve moving all of the joints at once. The goal of using MPC for a system such as this is to use model information to move all of the joints quickly and simultaneosly while reducing the overshoot and oscillation which are characteristic of under-damped systems. Each MPC controller is using a cost function of the form found in Equation \ref{eq:mpc_problem}, however the relative weightings on state and input error are tuned individually for each controller. Each controller used a horizon of 100 time steps and a discretization of .01 seconds, leading to a look-ahead time of one second.

MPC was run on one computer equipped with an intel E5-1603 CPU and Nvidia Titan X GPU while impedance control was run on a separate computer on the same local network.

%Because the model used for control is not very accurate, using the controllers by themselves leaves steady state error. This is an expected result of using MPC with an inaccurate model. While this could be improved with a better model, both the kinematic and dynamic modeling of soft robots is very difficult and is still an active area of research [cite things?]. 

%\todo{Do we use an integrator or not? If we do - explain it here. If not - explain that we should look more at the dynamic performance.}

\subsection{Hardware Experiments - Results}

\begin{figure*}[hbt]
  \centering
  \includegraphics[width=1.0\textwidth]{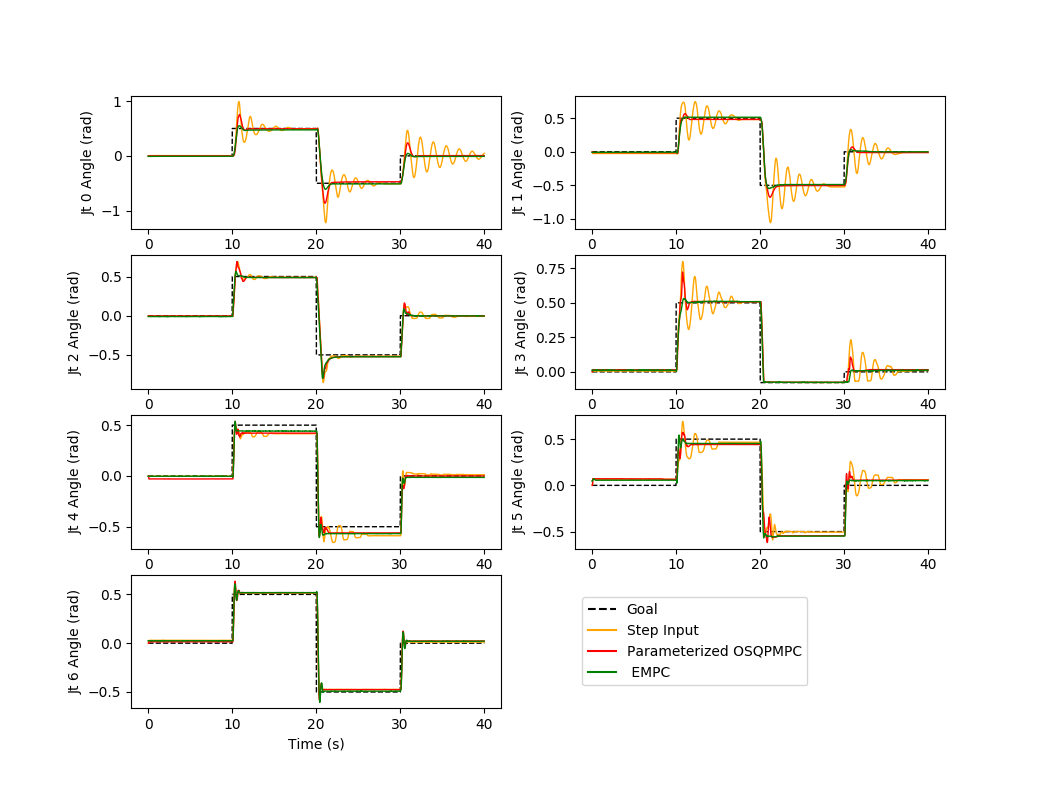}
  \caption{Joint angle response for the seven joints of the robot during the hardware experiment.}
  \label{fig:baxter_jangles}
\end{figure*}

As can be seen in Figure \ref{fig:baxter_jangles}, both MPC controllers are able to control the robot to the commanded joint positions with little to no overshoot or oscillation. In order to demonstrate the natural underdamped-ness of the system, the step response is also included in Figure \ref{fig:baxter_jangles}. While overshoot and oscillation can also be mitigated using command smoothing techniques, MPC allows for faster rise times than these methods afford.

The performance of the three control methods is quantified using the Integral Time Absolute Error (ITAE). This is defined as
\begin{equation}
    ITEA = \int_{t_0}^{t_1} (\tau-t_0)|q_{cmd}-q(t)| d\tau.
\end{equation}
This integral must be performed for each joint and each step input, resulting in 21 values for our experiment with seven joints and three step inputs. For easier comparison, we report the mean and median of these these 21 values for each trajectory in Table \ref{tab:ITAE}. As can be seen in the table, Parameterized OSQPMPC and EMPC reduce the median ITAE by 31\% and 42\% respectively.

The accuracy of the model leads to impressive MPC dynamic performance, however there is slight steady state error. This is due to the fact that the gravity compensation is not perfect. This could be decreased with an integrator or by using higher gains for the low level PD controller, however increasing the PD gains would lead to a stiffer system which is less safe in human environments.

\begin{table}[h!]
  \begin{center}
    \caption{Integral Time Absolute Error statistics of joint trajectories for hardware experiment}
    \label{tab:ITAE}
    \begin{tabular}{l|c|c}
      \textbf{Controller} & \textbf{Mean ITAE} & \textbf{Median ITAE} \\
      \hline
      %\textbf{Step Input} & 432.6 \\
      Step Input & 1.61 & 1.30\\
      \hline
      %\textbf{Parameterized OSQPMPC} & 313.2 \\
      Parameterized OSQPMPC & 1.29 & .90\\
      \hline
      %\textbf{LEMPC} & 293.3
      EMPC & 1.17 & .75
    \end{tabular}
  \end{center}
\end{table}

\begin{figure}[hbt]
  \centering
  \includegraphics[width=1.0\linewidth]{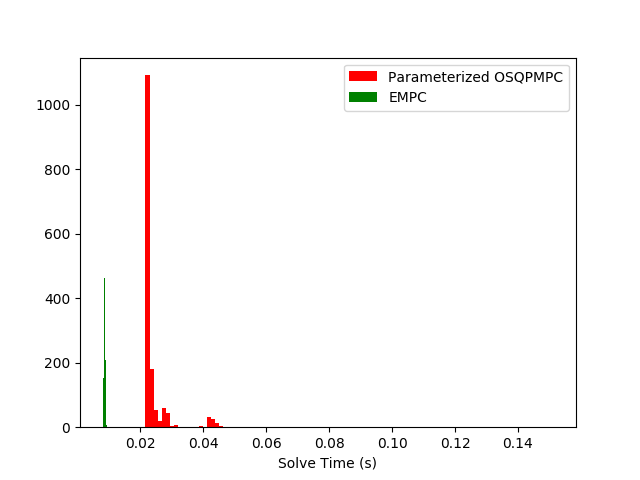}
  \caption{Hardware experiment solve times for the parameterized gradient-based MPC and EMPC}
  \label{fig:baxter_solve_times}
\end{figure}

While both controllers were able to successfully control this robot, they were not able to run at the same rate. Figure \ref{fig:baxter_solve_times} shows a histogram of MPC solve times for both EMPC and the parameterized gradient-based MPC. As can be seen from the figure, EMPC is able to solve consistently faster than the gradient-based method and also has far less variation in solve time. These hardware results demonstrate that by parallelizing the MPC problem, and offloading much of the computation to the GPU, MPC solve times can be greatly reduced without great loss of actual performance. This also means that the EMPC solver likely could have been run with a longer horizon if desired while still solving faster than the parameterized gradient-based method.

We should also note that this comparison does not include a traditional formulation of MPC because traditional MPC could not solve fast enough for real-time control given a horizon of 100 time steps. While EMPC enabled very fast MPC solves compared to parameterized gradient-based MPC, the fact that the gradient-based MPC is able to be run in real-time at 100 Hz is only due to the parameterization of the input trajectory that we present in this paper.

% \begin{figure}[hbt]
%   \centering
%   \includegraphics[width=1.0\linewidth]{figures/kaa_lempc_vs_osqp_inputs.png}
%   \caption{Inputs applied to the six joints of the robot during the hardware experiment. Inputs are goal pressures for a low level pressure controller. There are two pressurized chambers in each joint to be controlled.
%   \todo{I can re-take this data to smooth out joints 5 and 6. I should also make the colors match previous experiments}}
%   \label{fig:kaa_inputs}
% \end{figure} 

\section{Conclusion}\label{conclusion}

In this work we have shown for the presented form of input parameterization in the application of robot manipulators that input parameterization:
\begin{itemize}
    \item Produces more conservative trajectories, slightly favoring robustness over aggressive behavior
    \item Can be implemented using traditional convex solvers for very fast solve times
    \item Enables easily parallelized global optimization methods by drastically decreasing the optimization search space for high-DoF long horizon MPC problems
\end{itemize}
While we have shown these results only for robot manipulators, we expect that the methods presented in this paper may be generalized to other robotic systems.

% In this work we have addressed concerns related to performing MPC for high degree of freedom systems. We have presented a method of control trajectory parameterization which can be easily incorporated into the MPC framework and which is shown to decrease MPC solve times with little effect on performance for high degree of freedom systems. This parameterization does not decrease, but rather increases MPC robustness to modeling error by acting more conservatively.

The input-parameterized convex solver MPC presented in this paper was shown to drastically decrease solve times for high-DoF systems while maintaining long horizons. Many MPC controllers for robotics applications currently use this type of convex solver. We expect that most, if not all, may benefit from the input parameterization presented in this paper. The MPC controllers could either be made to run at higher rates using the same horizon length, or at the same rate using a longer horizon.

We have developed a parallelized form of MPC (EMPC) which can be run using a GPU and have shown that the solutions found using this method are also similar to that of traditional MPC. EMPC is shown to have faster solve times than both traditional MPC and the parameterized convex solver MPC for high DoF systems. Furthermore, experiments performed in this work demonstrate that EMPC solve times can be decreased by using a higher performance GPU. This indicates that with the development of better GPUs, parallelized MPC methods could be run at faster rates, with longer horizons, or could find higher quality solutions.

Another advantage of the EMPC approach which was not explored in detail in this work is the ability to change the cost function or dynamics into a form not previously admitted by MPC solvers. Because the optimization is gradient-free, it is possible to include sharp discontinuities in the cost if desired. Because system states are simulated over the horizon, the cost function could even be a direct function of rise time, settling time and overshoot. Furthermore, the assumption of linear dynamics could be relaxed and nonlinear models could be simulated on the GPU. Although this would lead to many local minima instead of one global minimum, it would more accurately represent a real system and because EMPC employs a global gradient-free optimization, a global minimum could be found.

A current weakness of the EMPC approach, and any sampling-based optimization used for MPC, is that it would be very difficult to prove stability using this controller. In practice we find that because we are able to sample such a large number of trajectories, we are able to find ``good" trajectories fairly quickly. Also, by making the exploration noise ($\sigma_{noise})$ a function of distance to a goal state, we find that once near the goal, EMPC keeps the system near the goal.
It seems that, if desired, statistical likelihood of stability using a sampling-based MPC could be calculated, however a guarantee seems impossible.

In this work we have only explored one form of parameterization (piecewise linear functions). This parameterization is linear and so was able to be fit into the form required by gradient-based solvers, however many different parameterizations are possible using the EMPC approach. Future work may include exploring parameterizations such as combinations of learned basis functions or other parameterizations of open-loop trajectories. 

If a parameterization is chosen for a feedback gain matrix it may even be possible to use a parallelized MPC solver such as EMPC to search the space of feedback controllers or policies directly. This could potentially lead to an MPC solver which returns a feedback controller which can be run at a high rate to minimize a given cost function over a finite horizon. While this idea is similar to iLQR or DDP, the EMPC based approach enables the global gradient-free search of a nonlinear space which may find better solutions than the gradient-based search used in iLQR and DDP.

Future work may also include other parallelizable optimization methods apart from the evolutionary strategy used in this work and implementations to other robotic platforms. The results presented in this work provide evidence that through the use of parameterization and parallelization, MPC may be applied to many high DoF systems for which it was not previously tractable.

\section*{Acknowledgements}
This work was partially funded by NSF Emerging Frontiers in Research and Innovation grant \# 1935312.

\bibliography{references_no_url}
\bibliographystyle{IEEEtran}

\end{document}